\pdfoutput=1
\documentclass[11pt]{article}
\usepackage{acl}
\usepackage{times}
\usepackage{latexsym}
\usepackage[T1]{fontenc}
\usepackage[utf8]{inputenc}
\usepackage{microtype}
\usepackage{inconsolata}
\usepackage{graphicx}
\usepackage{kotex}
\usepackage{xcolor}
\usepackage[normalem]{ulem}
\usepackage{xcolor}
\usepackage{color}
\usepackage{rotating}
\usepackage{amsmath}
\usepackage{array}
\usepackage{enumitem}
\usepackage[linesnumbered,algoruled,boxed,lined]{algorithm2e}
\usepackage{kotex}
\usepackage{balance}
\usepackage[short]{optidef}
\usepackage{graphicx}
\usepackage{tcolorbox}
\usepackage{hyperref}
\usepackage{multirow}
\usepackage{booktabs}
\usepackage{amsfonts}
\usepackage{tabularx}

\newcommand{\ie}{{\it i.e.}}
\newcommand{\eg}{{\it e.g.}}
\newcommand{\ul}{\underline}{}
\newcommand{\ours}{C{\small ORAL}}
\newcommand{\pearl}{P{\small EARL}}
\newcommand{\inspired}{I{\small NSPIRED}}
\newcommand{\redial}{R{\small E}D{\small IAL}}
\newcommand{\aours}{C{\footnotesize ORAL}}

\newcommand{\sours}{C{\scriptsize ORAL}}
\newcommand{\spearl}{P{\scriptsize EARL}}
\newcommand{\sinspired}{I{\scriptsize NSPIRED}}
\newcommand{\sredial}{R{\scriptsize E}D{\scriptsize IAL}}

\title{Empowering Retrieval-based Conversational Recommendation with Contrasting User Preferences}

\author{
    Heejin Kook\thanks{Equal contribution.}, Junyoung Kim\footnotemark[1], Seongmin Park, Jongwuk Lee\thanks{Corresponding author.} \\
    Sungkyunkwan University, Republic of Korea \\
    \texttt{\{hjkook, junyoung44, psm1206, jongwuklee\}@skku.edu}
}

\begin{document}
\maketitle
\begin{abstract}

Conversational recommender systems (CRSs) are designed to suggest the target item that the user is likely to prefer through multi-turn conversations. Recent studies stress that capturing sentiments in user conversations improves recommendation accuracy. However, they employ a single user representation, which may fail to distinguish between contrasting user intentions, such as \textit{likes and dislikes}, potentially leading to suboptimal performance. To this end, we propose a novel conversational recommender model, called \emph{\textbf{CO}ntrasting user p\textbf{R}eference exp\textbf{A}nsion and \textbf{L}earning (\textbf{\ours}}). Firstly, \aours\ extracts the user's hidden preferences through \emph{contrasting preference expansion} using the reasoning capacity of the LLMs.
Based on the potential preference, \aours\ explicitly differentiates the contrasting preferences and leverages them into the recommendation process via \emph{preference-aware learning}. Extensive experiments show that \aours\ significantly outperforms existing methods in three benchmark datasets, improving up to 99.72\% in Recall@10. The code and datasets are available at \url{https://github.com/kookeej/CORAL}.
\end{abstract}

% With the significant progress in large language models (LLMs), recent studies have explored using LLMs as recommenders or agents in CRSs.
% However, a great deal of existing work still fails to adequately address ambivalent user preferences, such as like and dislike, which may involve conflicting intentions. it leads to suboptimal performance.
% However, existing works overlook ambivalent user preferences, such as \textit{likes and dislikes} with conflicting intentions, leading to suboptimal performance.
% 고려하지만 보면 안된다.
% By utilizing LLMs to infer implicit user preferences, \ours\ differentiates the distinction between like/dislike preferences and the retrieval process. Through preference-aware learning, \ours\ ensures that LLMs capture subtle user preference similarities. Extensive experiments show that \ours\ significantly outperforms existing methods across all datasets.
% The expanded datasets and source code will be available upon acceptance.
\section{Introduction}
\label{sec:introduction}

%% 1. CRS 설명, 두 가지 Task 설명
%% 2. Traditional CRS 모델 소개 및 한계점 / Preference-aware CRS 모델 언급
%% 3. Existing Preference-aware CRS 모델 소개 및 한계점
%% 4. 두 가지 Question 던지기
%%    4-1. Q1: 사용자가 좋아하고 싫어하는걸 어떻게 알거고, 이걸 어떻게 표현할거임? (how to represent)
%%    4-2. Q2: 사용자가 좋아하고 싫어하는 걸 어떻게 item과 align해서 추천할거임? (how to align)
%% 5. 제안 방법 소개
%%    5-1. A1: 추출하고 증강하고 text로 표현 (semantic, context 키워드 강점)
%%    5-2. A2: representation learning (user <-> positive preference and negative preference <-> item)
%%         - retrieval-based CRSs의 이점 언급
%% 6. Contribution

%% 1. CRS 설명
%Recent advancements in natural language understanding have accelerated the development of Conversational Recommender Systems (CRSs). 
Conversational recommender systems (CRSs)~\cite{pref/crs2021revcore, kg/crs2022unicrs, pref/crs2024ecr, llm/crs2024chatcrs} deliver personalized recommendations by deeply understanding users' evolving contexts through multi-turn interactions. Typically, CRSs consist of the following two tasks: \emph{recommendation task} to provide items by identifying the user's intent from the text, and \emph{generation task} to offer a human-friendly response to the user. While recent LLMs have shown impressive performance in natural language generation~\cite{llm/crs2023llmzs, agent/crs2024interecagent}, the recommendation task is yet challenging to address. In this paper, we thus focus on improving the recommendation task.

\begin{figure}[t]
\centering
\includegraphics[width=1.0\linewidth]{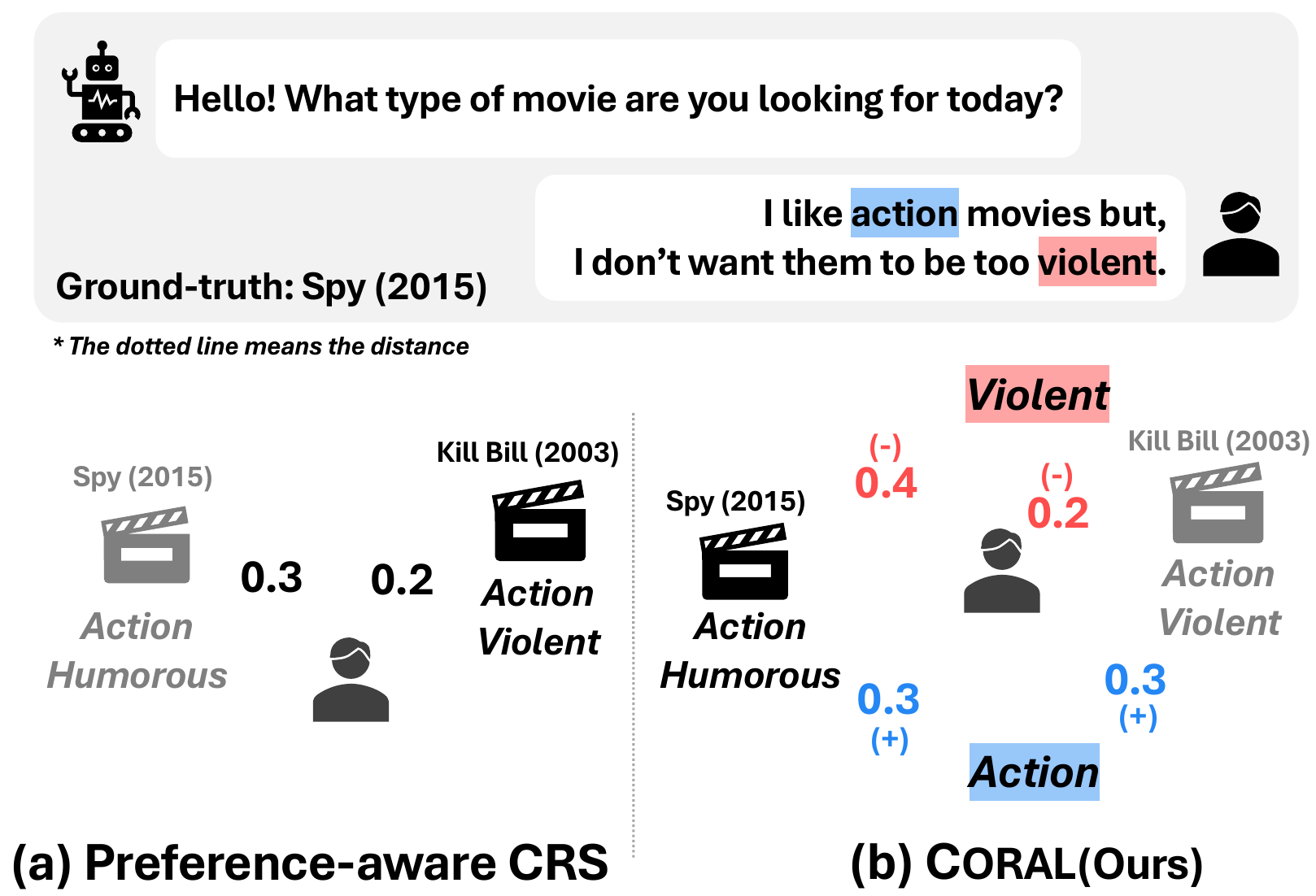} 
\caption{Comparison between traditional preference modeling and contrasting preference-aware modeling.}
\label{fig:motivating}
\end{figure}

%% 2. 기존 CRS 모델 소개 및 한계점 / Preference-aware CRS 모델 언급
% 한계점: (일단 ECR 한계점으로 적어보기)
Existing CRS methods~\cite{kg/crs2019kbrd, kg/crs2020kgsf, kg/crs2022unicrs} leverage external information (\eg, knowledge graphs) or LLMs' reasoning capabilities to understand dialogue context and recommend user's preferable items. 
While user conversations often contain both positive and negative user preferences, they assume that the entities in the dialogue history are positive.
Capturing various user sentiments or emotions is critical for understanding hidden user preferences in the decision-making process~\cite{lerner2015emotion, pref/crs2024ecr}.
% Existing CRS approaches~\cite{kg/crs2019kbrd, kg/crs2020kgsf, kg/crs2022unicrs}은 external information (\eg, knowledge graph)나 LLMs의 reasoning 능력을 활용하여 dialogue context를 이해하고, user가 선호할만한 items을 추천한다.
% 이들은 conversation에 언급된 entity들이 positive하다고만 여긴다. 하지만, user conversation에는 positive와 negative 의미가 혼재하며, human decision-making 과정에서 user sentiment/emotion를 포착하는 것은 user preference modeling함에 있어 중요하다~\cite{lerner2015emotion, pref/crs2024ecr}.

%% 3. Existing Preference-aware CRS 모델 소개 및 한계점 (Motivated Figure)
To address this issue, recent studies~\cite{pref/crs2021revcore, pref/crs2024ecr, llm/crs2024memocrs, rebuttal2023FPAN, rebuttal2023MHCPL} introduce preference-aware CRS models that extract user preferences from conversations across multiple sentiments or emotions, leveraging them for more precise item recommendations. For example, \citet{pref/crs2024ecr} estimated probabilities for nine preference labels for each user utterance and integrated these probabilities into a unified user representation.
However, they failed to model the complicated relationship among the user, items, and individual preferences by representing contrasting preferences with a single user representation.

% This highlights the need to (1) differentiate between different preferences and (2) learn not only the user-preference relationship but also the user-preference-item relationship.

%It fails to differentiate and consider ambivalent preferences may lead to undesirable results.
% 대화에서 사용자는 서로 반대되는 preference를 나타낸다 (namely ambivalent preference)

% Challenges
In the conversation, users express various preferences with opposing intentions, such as like and dislike, which can be referred to as \emph{contrasting preferences}. Figure~\ref{fig:motivating} illustrates an example of the importance of differentiating contrasting preferences, where the number indicates the semantic distance between the user and the item.
% 표시된 숫자는 사용자와 아이템 사이의 거리를 의미한다.
Although the user expresses a negative sentiment towards \emph{``violent''}, existing preference-aware CRS models may still recommend \emph{``Kill Bill (2003)''}, which contains a similar characteristic to the user utterance. In contrast, considering negative preferences directly enables us to recommend \emph{``Spy (2015)''}, where the user may exhibit less negative sentiment toward \emph{``violent''}.

%% 4. 두 가지 Question 던지기
%%    4-1. Q1: 사용자가 좋아하고 싫어하는걸 어떻게 알거고, 이걸 어떻게 표현할거임?
%%    4-2. Q2: 사용자가 좋아하고 싫어하는 걸 어떻게 item과 align해서 추천할거임?
Based on this observation, we ask the following key questions: \textit{(i) How do we extract contrasting user preferences from the conversation? (ii) How do we learn the relationship between the contrasting preferences and the user/item?}

%% 5. 제안 방법 소개
%%    5-1. A1: 추출하고 증강하고 text로 표현 (semantic, context 키워드 강점)
%%    5-2. A2: representation learning (user <-> positive preference and negative preference <-> item)
%%         - retrieval-based CRSs의 이점 언급 (Motivated Figure)
To this end, we propose a novel retrieval-based CRS framework, \emph{\textbf{CO}ntrasting user p\textbf{R}eference exp\textbf{A}nsion and \textbf{L}earning (\textbf{\ours})}, which extracts and learns contrasting preferences. Specifically, it has two key components. Firstly, we utilize \emph{contrasting preference expansion} using the advanced reasoning power of LLMs to accurately distinguish contrasting preferences within user-system dialogues into positive (\ie, \emph{like}) and negative (\ie, \emph{dislike}) preferences. Then, the extracted preferences are augmented for the recommendation task to elicit the user's potential preferences. In this process, we utilize a dense retrieval model to extract users, items, and preferences, thus representing them within the same representation space.

% retriever 방식 활용의 이점은 다음과 같다: (i) 모두 같은 모달리티와 모델을 공유해서 사용자, 아이템, preference 모두 동일한 representation space에 표현할 수 있고,

% Firstly, \emph{Preference extraction module}은 LLMs의 뛰어난 reasoning ability를 활용하여 user와 system 사이의 대화 내 존재하는 암시적인 preference를 positive preference (\ie, \emph{like})와 negative preference (\ie, \emph{dislike})로 정확하게 구분해 추출한다. 그 다음, 추출한 preference를 recommendation task에 맞게 증강하여 사용자의 potential preference를 추론한다.

% Secondly, \emph{preference-aware learning module}은 상반된 두 가지의 potential preference를 매개체로 활용하여 recommenders의 효과적인 학습을 수행한다, as depicted in Figure~\ref{fig:motivating}(b). 
% 우리는 먼저 서로 다른 특징을 가진 dialogues, preference, item 간의 관계를 모델링하기 위해 이 세 정보를 모두 text 형태로 표현한 후, dense retriever 구조를 사용하여 모델링한다.
% Given the textual representations (\ie, ambivalent preferences, dialogue, and items) obtained from the ambivalent preference expansion stage, we model their relationships using a dense retrieval structure.

Secondly, \emph{preference-aware learning} is used to capture two opposite user preferences to identify whether the user will like or dislike the given item, as depicted in Figure~\ref{fig:motivating}(b). Given the textual representations, which consist of dialogue and item descriptions, along with like/dislike preferences obtained from the contrasting preference expansion stage, we input them into an encoder and optimize item representations to be semantically closer to dialogue and like preferences while being pushed further apart from dislike preferences. We also use negative sampling to enable the model to distinguish between items that are difficult to classify based solely on conversation representation, enhancing preference representation. Therefore, it allows us to directly associate both preference types to calculate the recommendation scores of items.

% ambivalent preference expansion에서 얻은 text로 표현된 preference와 dialogue와 item가 주어지면, 우리는 그들 간의 관계를 dense retrieval 구조를 사용하여 모델링한다.
% 그 후, \emph{preference modeling}을 통해 positive preference와는 user와 item을 가깝게 학습하고, negative preference와는 user와 item을 멀어지게 학습한다.
% 또한, conversation을 기반으로 hard negative sampling을 통해 preference를 leverage할 수 있도록 한다.
% 이를 통해, recommended item ranking/scoring에 직접적으로 두 가지 유형의 preference를 반영할 수 있다.
% retriever 방식 활용의 이점은 다음과 같다: (i) 모두 같은 모달리티와 모델을 공유해서 같은 representation space에 표현할 수 있고, (ii) 사용자가 같은 preference를 보이는 aspects을 LLMs에 통과시켜 general하고 contextual한 표현을 얻을 수 있다.

% (ii) 언어 모델의 지식을 활용하여 preferences를 context에 맞게 그리고 general knowledge를 함의하도록 표현할 수 있다.
% xx할 수 있다.
% Furthermore, 우리는 recommender로 retriever를 사용하기 때문에, alignment learning을 하기에 적합하다.

The main contributions of our work are summarized as follows:
\begin{itemize}[leftmargin=5mm]
    \item \textbf{Recommendation-tailored Augmentation}: We extract contrasting user preferences to achieve effective user preference modeling. Leveraging the reasoning capabilities of LLMs, we identify complex preferences expressed in natural language within the dialogue and augment potential preferences using prompts tailored for recommendation tasks.
    \item \textbf{Preference-aware Recommender}: To directly involve contrasting user preferences, we explicitly model and learn the relationships among the users, preferences, and items. Explicitly separating both like and dislike preferences provides a rationale for the recommendations, enhancing the interpretability and transparency of our approach.
    
    \item \textbf{Comprehensive Validation}: \ours~outperforms seven baselines across three datasets, improving up to 99.72\% in Recall@10. Notably, the ablation study demonstrates the effectiveness of learning preference relationships by separating like/dislike from user preferences.
    % \ours는 다양한 preference sparsity를 갖는 세 가지의 datasets에서 4종류의  6개의 모델들을 outperform한다. 특히 ablation study를 통해 like/dislike를 아이템과 분리하여 preference relation을 학습하는 것이 효과적임을 보인다.

\end{itemize}

\section{Preliminaries} \label{sec:preliminary}

\subsection{Problem Statement}
Let $u$ and $i$ denote a user and an item of user set $\mathcal{U}$ and item set $\mathcal{I}$. Each item $i$ contains a key-value list of metadata represented as $\{(a_m, v_m)\}_{m=1}^{|i|}$, where $a_m$ and $v_m$ denote the textual attribute (\eg, Title) and the corresponding textual value (\eg, Frozen(2013)) of $m$-th metadata, respectively. Here, $|i|$ represents the number of metadata entries associated with item $i$.
The dialogue history of $u$ is denoted as $c=\{(s_t, u_t)\}_{t=1}^{|c|}$, where $u_t$ is the utterance at $t$-th turn, $|c|$ is the number of turns within $c$ and $s_t$ is the speaker at $t$-th turn, either the \emph{user}, seeking an item or the \emph{system}, providing personalized recommendations, respectively.
% where $s_t$ and $u_t$ are speaker and utterance at $t$-th turn, $|c|$ is the number of turns within $c$.
% Here, $s_t$ denotes either the \emph{user}, seeking to find a desired item, or the \emph{system}, which identifies user preferences to provide personalized recommendations.

The goal of CRSs is to offer a set of candidate items to the user at the $n$-th turn, based on the dialogue history $c=\{(s_t, u_t)\}_{t=1}^{n-1}$ and the available metadata.

\begin{figure*}[t]
\centering
\includegraphics[width=1.0\linewidth]{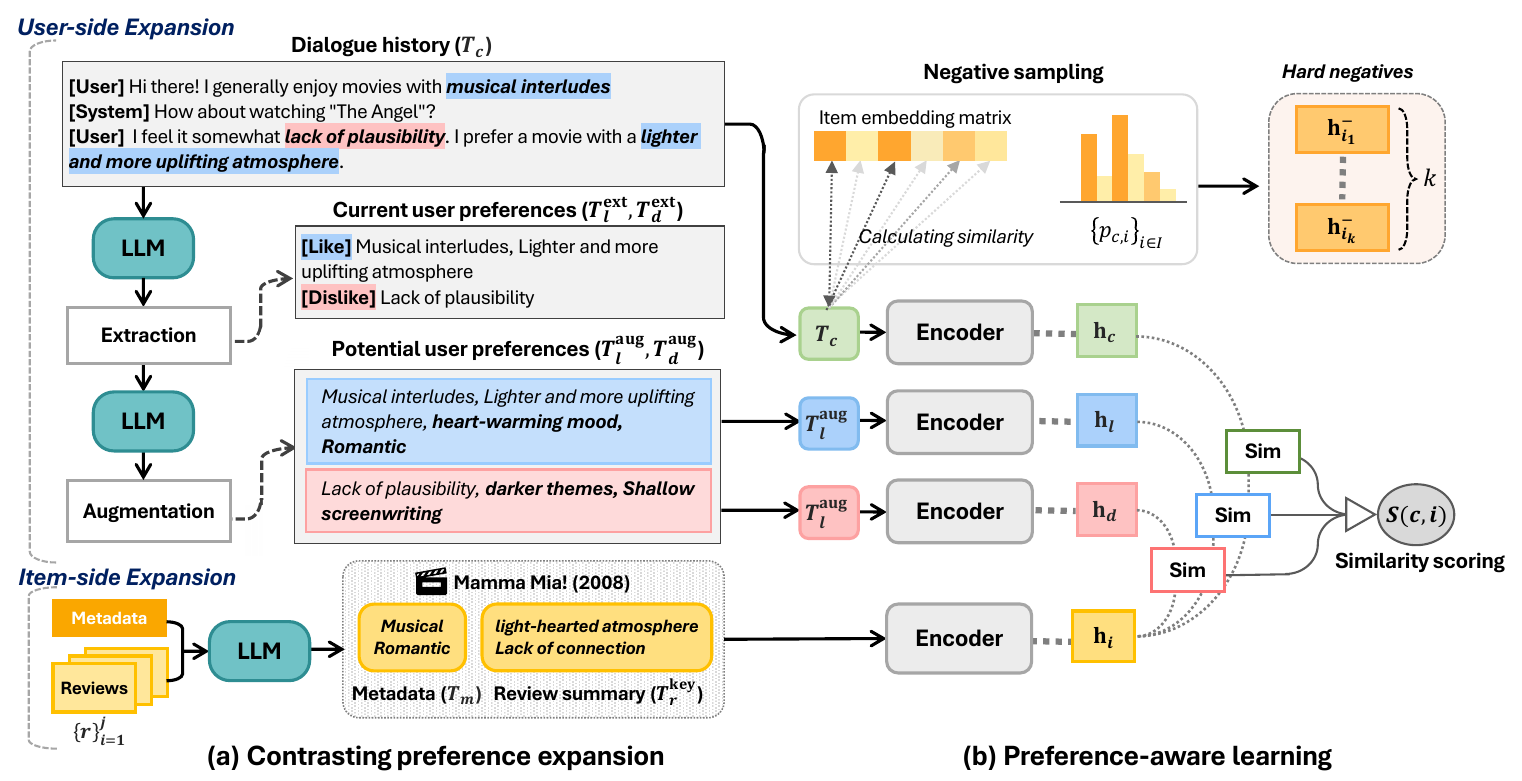} 
\caption{Overall architecture of \ours. It comprises two components: (i) \emph{Contrasting Preference Expansion}, which extracts superficial user preferences and augments potential preferences implicitly present in the conversation (Section~\ref{sec:contrasting_preference_expansion}); and (ii) \emph{Preference-aware Learning}, which directly models the relationships among the user, contrasting preferences, and items (Section~\ref{sec:preference_learning}).}
\label{fig:main}
\end{figure*}

\subsection{Retrieval-based CRSs}
Retrieval-based CRS models~\cite{ret/crs2023CRSRetbaseline} recommend items based on the similarity between the textual representations of the dialogue $T_c$ and the item $T_m$. Specifically, $T_c$ is obtained by concatenating the speaker and utterance of every turn in the conversation (\eg, $T_c$ = \textit{``[User] Hi there! ... [System] How about ...''}).
The similarity score $\mathrm{sim}(c, i)$ between $c$ and $i$ is calculated as follows:
\begin{equation} \label{overall_scoring}
    S(c,i) = \mathrm{sim}(\mathrm{Enc}(T_c), \mathrm{Enc}(T_m)),
\end{equation}
where $\mathrm{sim}$ denotes the similarity function (\eg, dot product), and $ \mathrm{Enc}$ denotes the encoder-based language models such as BERT~\cite{lm2019bert}.

\section{Proposed Model: \ours} \label{sec:Proposed Method}
This section presents a novel CRS framework \ours, as depicted in Figure~\ref{fig:main}, designed to differentiate between contrasting user preferences and effectively associate them with items.

\subsection{Contrasting Preference Expansion}
\label{sec:contrasting_preference_expansion}
Figure ~\ref{fig:main}(a) illustrates the overall process of contrasting preference expansion. We perform the user- and the item-side preference expansion. The user-side expansion aims to distinguish and enhance contrasting preferences into like/dislike ones, and the item-side expansion addresses the discrepancy between dialogue and item metadata.
% 우리는 user-side와 item-side에 대해서 모두 preference expansion을 수행한다.

\subsubsection{User-side Expansion}
% We utilize an LLM to analyze the dialogue history and accurately extract the user's superficial preferences in the form of keyphrases. These keyphrases serve as evidential input, enabling the LLM to infer additional implicit and potential preferences with the conversation~. This two stage approach enhacnes the relevance and accuracy of recomendations by incorporating both superficial and potential preferences.
User-side expansion distinguishes and infers user preferences embedded in the dialogue. We decompose the problem into two sub-tasks.
First, we utilize an LLM to analyze the dialogue history and accurately extract the user's superficial preferences. These preferences serve as evidential input, enabling the LLM to infer additional implicit and potential preferences from the conversation~\cite{llm2024SelfEvolvingGPT,llm2023selfrefine}.
% This two-stage process enhances the relevance and accuracy of recommendations by incorporating both superficial and potential preferences.

\vspace{0.5mm}
\noindent
\textbf{\texttt{Step 1}: Superficial preference extraction.}
For a given user $u$, we incorporate the dialogue into a prompt template $P_{\mathrm{ext}}$ and input it into LLMs. We provide a detailed example in the Appendix~\ref{sec:appendix_prompt}.
\begin{equation}
    T_{\mathit{l}}^{\mathrm{ext}}, T_{\mathit{d}}^{\mathrm{ext}} = f_{\mathrm{LLM}}(P_{\mathrm{ext}}, T_c),
\end{equation}
where $T_{\mathit{l}}^{\mathrm{ext}}$ and $T_{\mathit{d}}^{\mathrm{ext}}$ represent lists of keyphrases that capture the like and dislike preferences of user $u$, respectively. By leveraging the reasoning power of LLM, it is possible to analyze the context of dialogue history and accurately extract the user's meaningful surface preferences, allowing for precise distinction between likes and dislikes.

\vspace{0.5mm}
\noindent \textbf{\texttt{Step 2}: Potential preference augmentation.}
To improve recommendation accuracy, we leverage the reasoning power of LLMs to augment the user's potential preferences.
We use the extracted superficial preferences of user $u$ as a rationale to infer potential preferences that the user might like,  following self-feedback approaches in previous studies~\cite{llm2023selfrefine, llm2024SelfEvolvingGPT}.
\begin{equation}
    T_{\mathit{l}}^{\mathrm{aug}}, T_{\mathit{d}}^{\mathrm{aug}} = f_{\mathrm{LLM}}(P_{\mathrm{aug}},T_c, T_{\mathit{l}}^{\mathrm{ext}}, T_{\mathit{d}}^{\mathrm{ext}}),
\end{equation}
where $ T_{\mathit{l}}^{\mathrm{aug}} $ and  $T_{\mathit{d}}^{\mathrm{aug}}$ denote lists of keyphrases that infer the potential positive and negative preferences of user $u$, respectively.

\subsubsection{Item-side Expansion}
\label{sec:item-side expansion}
% Item reviews reflect the common preferences of many users and can be leveraged to enhance the expressiveness of items, improving search performance in dense retrieval-based approaches. Specifically, the item metadata typically used in dense retrieval may not sufficiently capture user preferences, potentially creating a gap between user conversations and items, which can be bridged by the rich information in reviews.
To enhance the representation of an item, we adopted review summaries.
Item metadata often diverges from the nature of conversations due to the absence of preferences.
The preferences in the review bridge the gap between user conversation and item metadata, making the item representations more expressive.
% Common user preferences reflected in the reviews improve search performance in retrieval-based approaches by capturing nuances often missed in item metadata.
% 리뷰에는 여러 사용자의 선호 정보가 포함되어 있어, 사용자의 선호를 반영하는 대화와의 매칭에 효과적이다. 반면, 아이템의 메타데이터는 이러한 대화의 본질적 특징과 다소 차이가 있다. 따라서, 메타데이터에 포함되지 않은 선호 정보를 추가하여, 대화와 메타데이터 간의 연결을 돕는 일종의 브릿지 역할을 수행할 필요가 있다.
% Reviews capture essential user preferences, making them effective for matching preference-driven conversations.
% However, item metadata often diverges from the nature of conversations.
% Thus, incorporating missing preference information is crucial in bridging the gap between dialogues and metadata.

% While users' potential preferences are augmented through LLMs,
% Review summaries can help enhance item expressiveness by bridging the gap between user conversations and item representation. Item reviews, reflecting common user preferences, improve search performance in retrieval-based approaches by capturing nuances often missed in item metadata.

Specifically, we crawled reviews for item $i$ from IMDb\footnote{https://www.imdb.com/} and selected the top-$j$ reviews based on helpful votes. Using an LLM, we extracted and summarized common preferences from the reviews and identified keyphrases to optimize item expressiveness. This process condenses the representation, diminishes redundancy, and enhances contextual relevance.
\begin{equation}
    T_{\mathit{r}}^{\mathrm{sum}} = f_{\mathrm{LLM}}(P_{\mathrm{summary}}, T_{\mathrm{m}}, \{r\}_{i=1}^{j}),
\end{equation}
where $r$ denotes individual reviews that have been crawled, and the top-$j$ reviews are inserted into the prompt $P_{\mathrm{summary}}$, which is designed to summarize multiple reviews. The common preferences summary for all the generated reviews is $T_{\mathit{r}}^{\mathrm{sum}}$, which is then processed through an LLM to extract keyphrases using prompt template $P_{\mathrm{keyphrase}}$.
\begin{equation}
    T_{\mathit{r}}^{\mathrm{key}} = f_{\mathrm{LLM}}(P_{\mathrm{keyphrase}}, T_{\mathit{r}}^{\mathrm{sum}}).
\end{equation}

Further explanation, along with concrete examples of this process, is provided in Appendix~\ref{sec:appendix_prompt}.

\subsection{Preference-aware Learning}
\label{sec:preference_learning}
Existing studies~\cite{pref/crs2021revcore, pref/crs2024ecr} utilize differentiated user preferences but a single representation to represent contrasting preferences. In contrast, we explicitly represent these preferences separately from the conversation and learn the relationship between them to engage the preferences in item scoring directly.

\subsubsection{Preference Modeling}
\label{sec:matching}
\textbf{Representation.}
The encoding function $f_{\text{Enc}}$ takes the given dialogue history $T_c$ as input and returns a vector representation of the dialogue history $c$, denoted as $\mathbf{h}_c = f_{\text{Enc}} (T_c) \in \mathbb{R}^d$.
% The encoding function $f_{\text{Enc}}$ and the dialogue history $T_c$, the vector representation of the dialogue history $c$ is denoted as $\mathbf{h}_c = f_{\text{Enc}} (T_c) \in \mathbb{R}^d$.
Specifically, $T_c$ is tokenized and passed through a text encoder($\text{Enc}$). The last hidden states of all tokens are subsequently mean-pooled, followed by L2 normalization to derive $\mathbf{h}_c$.
\begin{align}
    % h_c = f_\mathrm{Enc}(T_c),\text{where} \\
    [t_1, \dots, t_w] & = \text{Tokenize}(T_c) \\
    [\mathbf{o}_1, \dots, \mathbf{o}_w] & = \text{Enc}([t_1, \dots, t_w]) \\
    \mathbf{h}_c = \frac{\mathbf{o}_c}{\|\mathbf{o}_c\|_2},&~\text{where}~\mathbf{o}_c = \frac{1}{w} \sum_{i=1}^{w} \mathbf{o}_i,
\end{align}
where $t$ is the token, $w$ is the length of tokenized $T_c$ and $\mathbf{o}_i \in \mathbb{R}^d$ is the vector of last hidden state.

% Similarly, for $T_{\mathit{l}}^{\mathrm{aug}}$ and $T_{\mathit{d}}^{\mathrm{aug}}$, we obtain the two ambivalent preference representations, $\mathbf{h}_l$ and $\mathbf{h}_d$.
Similarly, for $T_{\mathit{l}}^{\mathrm{aug}}$ and $T_{\mathit{d}}^{\mathrm{aug}}$, we obtain the like/dislike preference representations, $\mathbf{h}_l$ and $\mathbf{h}_d$.
\begin{align}
    \mathbf{h}_l = f_\mathrm{Enc}(T_{\mathit{l}}^{\mathrm{aug}})~\text{and}~\mathbf{h}_d = f_\mathrm{Enc}(T_{\mathit{d}}^{\mathrm{aug}}).
\end{align}

Inspired by previous work~\cite{ret/rs2024blair, data2024AlignLM2IR}, we concatenate item metadata and review information to obtain a review-enhanced item vector representation.
\begin{align}
    \mathbf{h}_i &= f_\mathrm{Enc}(T_m \oplus T_{\mathit{r}}^{\mathrm{key}}),
\end{align}
where the $\oplus$ denotes concatenation.

Consequently, we obtain three user-side representations (\ie, $\mathbf{h}_c$, $\mathbf{h}_l$, and $\mathbf{h}_d$) and one item-side representation (\ie, $\mathbf{h}_i$).

\vspace{0.5mm}
\noindent \textbf{Similarity scoring.}
To compute the final score, we linearly aggregate three scores—the similarity between the dialogue history and item, between like preference and item, and between dislike preference and item. The goal is to ensure that the desired item is close to the conversation context and the user's like preference while being far from the user's dislike preference.
We extend Equation~\eqref{overall_scoring} as follows:
\begin{align} \label{eq:similarity_scoring}
S(c, i) &= \text{sim}(\mathbf{h}_c, \mathbf{h}_i) \nonumber \\
        &\quad + \alpha \cdot \, \text{sim}(\mathbf{h}_l, \mathbf{h}_i) - \beta \cdot \, \text{sim}(\mathbf{h}_d, \mathbf{h}_i),
\end{align}
where $\alpha, \beta \in (0, 1]$ are hyperparameters representing the importance of the user's like and dislike preferences. Empirically, $\alpha$ is set to 0.5, and $\beta$ is adjusted in [0.1, 0.3] depending on the dataset.

This approach is a simple yet effective way of reflecting contrasting preferences, requiring no additional parameters.
Furthermore, it enhances interpretability by intuitively revealing which preferences the recommended items are derived from, demonstrated in a case study (Section~\ref{sec:case_study}).
% 또한 이러한 방식은 추천된 item이 어떤 preference로부터 기인하는지 직관적으로 보여주며 해석 가능성을 높인다.

% 이는 추가로 파라미터를 필요로 하지 않는 간단한 방식이면서 ambivalent한 preference를 반영하는 효과적인 방식이다.

\subsubsection{Training}
\label{sec:training}
% Finally, 우리는 \ours를 학습하는 방법에 대해 설명한다.

\noindent \textbf{Hard negative sampling.}
Hard negative sampling directly impacts the convergence and performance of dense retrieval models~\cite{neg2021ANCE}.
% It is crucial in representation learning because it enhances the model's discriminative ability by focusing on difficult-to-distinguish samples, leading to more efficient learning~\cite{neg2021ANCE,neg2021CLw/Neg}.
% Additionally, the informativeness of these hard negatives directly impacts the convergence and performance of dense retrieval models.

Our key contribution lies in utilizing hard negative sampling to enhance the representation of user preferences, especially for samples that are challenging to predict based on conversation alone.
Specifically, we first compute the similarity between $\mathbf{h}_c$ and all item embeddings, then apply softmax to convert the similarity scores into a probability distribution. Using this distribution $\{p_{c,i}\}_{i \in I}$, we sample a set of $k$ negative items ${\mathcal{I}_c}^-$, resulting in a set of hard negative samples, further enriching the model's understanding of user preferences.
\begin{align}
    p_{c,i} &= \frac{\exp(\mathrm{sim}(\mathbf{h}_c, \mathbf{h}_i))}{\sum_{j \in \mathcal{I}} \exp((\mathrm{sim}(\mathbf{h}_c, \mathbf{h}_j))} \\
    {\mathcal{I}_c}^- &= \{i_1, i_2, \dots, i_k\} \nonumber \\
    & \sim \text{Multinomial}\left(k, \{p_{c,i}\}_{i \in \mathcal{I}}\right).
\end{align}

\noindent
\textbf{Loss function.}
For training, we utilize the cross-entropy loss $\mathcal{L}$ as follows.
\begin{equation} \label{eq:training_loss}
    \mathcal{L} = - \log\frac{\text{exp}\left(S \left( c,i^{+}\right) / \tau \right)}{\sum_{i \in {\mathcal{I}_c}^-} {\text{exp}\left( S\left( c,i \right) / \tau\right)}},
\end{equation}
where $i^{+}$ is the positive item of $c$, and $\tau$ is a hyperparameter to adjust the temperature.

\noindent
\textbf{Training procedure.}
For training efficiency, we compute $\mathbf{h}_c$ and $\mathbf{h}_i$ for all items $c$ and $i$ in the training set at the beginning of each epoch to obtain hard negative samples for the entire training set. The pre-computed $\mathbf{h}_i$ values are stored in an item embedding table, allowing the model to retrieve item embeddings directly from the table during training without passing them through the encoder again. This approach reduces the time complexity and enables more efficient training.
% For training efficiency, 우리는 epoch이 시작할 때 training set의 $c$와 $i$ 전체 아이템에 대해 $\mathbf{h}_c$와 $\mathbf{h}_i$를 계산하여 전체 training set에 대한 hard negative samples을 얻는다.
% 이때 계산한 전체 아이템의 $\mathbf{h}_i$를 item embedding table로써 활용하여 학습 시에는 item을 따로 encoder에 통과시키지 않고 item embedding table에서 가져온다.
% 이러한 방식은 학습의 시간 복잡도를 줄이고, 더 효율적인 학습을 가능하게 한다.

% \usepackage{booktabs}
% \usepackage{multirow}
\begin{table}[]\small
\centering
\setlength{\tabcolsep}{4pt}
\begin{tabular}{c|rrrr}
\toprule
\textbf{Dataset} & \textbf{\#Dial.} & \textbf{\#Items} & \textbf{\#Likes} & \textbf{\#Dislikes} \\ \midrule
\spearl    & 57,159 & 9,685 & 9.59  & 5.97 \\
\sinspired & 1,997  & 1,058 & 11.09 & 5.65 \\
\sredial   & 31,089 & 5,896 & 10.99 & 1.00 \\ \bottomrule
\end{tabular}
\caption{Data statistics. `Dial.' represents dialogue history, while `\# Likes' and `\# Dislikes' refer to the average counts of the like and dislike preferences after the augmentation stage, respectively.}\label{tab:data_statistics}
\end{table}

\iffalse
\begin{table}[]\small
\centering
\begin{tabular}{c|cccc}
\hline
\textbf{Dataset} & \textbf{\#Dial.} & \textbf{\#Items} & \textbf{Avg. Like} & \textbf{Avg. Dislike} \\ \hline
pearl    & 57,159 & 9,685 & 9.59  & 5.99 \\
inspired & 2,017  & 1,058 & 11.11 & 5.65 \\
redial   & 31,089 & 5,896 & 10.99 & 1.00 \\ \hline
\end{tabular}
\caption{Data statistics. \textit{Dial.} represents dialogue history, while \textit{Avg. likes} and \textit{Avg. dislikes} refer to the average length of the liked and disliked aspects, respectively.}\label{tab:data_statistics}
\end{table}
\fi
% \usepackage{multirow}
\begin{table*}[] \small
\centering
\begin{tabular}{cc|ccc|cc|ccc|c}
\toprule
\multicolumn{2}{c|}{Model} & \multicolumn{3}{c|}{Traditional CRS} & \multicolumn{2}{c|}{LLM-based CRS} & \multicolumn{3}{c|}{Retrieval-based CRS} & \multirow{2}{*}{Gain} \\ 
\multicolumn{1}{c|}{Dataset} & Metric & RevCore & UniCRS & ECR & Zero-shot & ChatCRS & BM25 & \multicolumn{1}{c|}{DPR} & \ours &  \\ \midrule
\multicolumn{1}{c|}{\multirow{4}{*}{\spearl}} & R@10 & 0.0268 & \ul{0.1156} & 0.0957 & 0.0767 & 0.0763 & 0.0026 & \multicolumn{1}{c|}{0.0940} & \textbf{0.1851*} & 60.07\% \\
\multicolumn{1}{c|}{} & R@50 & 0.0898 & \ul{0.2624} & 0.2373 & 0.1129 & 0.1168 & 0.0123 & \multicolumn{1}{c|}{0.2206} & \textbf{0.3619*} & 37.94\% \\ 
\multicolumn{1}{c|}{} & N@10 & 0.0132 & \ul{0.0642} & 0.0501 & 0.0468 & 0.0462 & 0.0014 & \multicolumn{1}{c|}{0.0502} & \textbf{0.1125*} & 75.17\% \\
\multicolumn{1}{c|}{} & N@50 & 0.0266 & \ul{0.0958} & 0.0806 & 0.0560 & 0.0565 & 0.0033 & \multicolumn{1}{c|}{0.0777} & \textbf{0.1511*} & 57.74\% \\ \midrule
\multicolumn{1}{c|}{\multirow{4}{*}{\sinspired}} & R@10 & 0.0948 & 0.1113 & \ul{0.1711} & 0.1436 & 0.1410 & 0.0429 & \multicolumn{1}{c|}{0.1019} & \textbf{0.3417*} & 99.72\% \\
\multicolumn{1}{c|}{} & R@50 & \ul{0.3344} & 0.2528 & 0.2826 & 0.2436 & 0.2436 & 0.1210 & \multicolumn{1}{c|}{0.2672} & \textbf{0.5632*} & 68.45\% \\ 
\multicolumn{1}{c|}{} & N@10 & 0.0509 & 0.0642 & \ul{0.1077} & 0.0927 & 0.0806 & 0.0202 & \multicolumn{1}{c|}{0.0512} & \textbf{0.1772*} & 64.52\% \\
\multicolumn{1}{c|}{} & N@50 & 0.1041 & 0.0952 & \ul{0.1417} & 0.1175 & 0.1071 & 0.0373 & \multicolumn{1}{c|}{0.0872} & \textbf{0.2255*} & 59.07\% \\ \midrule
\multicolumn{1}{c|}{\multirow{4}{*}{\sredial}} & R@10 & \ul{0.1739} & 0.1549 & 0.1685 & 0.1670 & 0.1666 & 0.0373 & \multicolumn{1}{c|}{0.0774} & \textbf{0.2182*} & 25.48\% \\
\multicolumn{1}{c|}{} & R@50 & 0.3034 & 0.3540 & \ul{0.3793} & 0.2783 & 0.2824 & 0.0300 & \multicolumn{1}{c|}{0.2138} & \textbf{0.4741*} & 25.23\% \\
\multicolumn{1}{c|}{} & N@10 & \ul{0.1053} & 0.0776 & 0.0805 & 0.0937 & 0.0893 & 0.0032 & \multicolumn{1}{c|}{0.0403} & \textbf{0.1128*} & 7.10\% \\
\multicolumn{1}{c|}{} & N@50 & \ul{0.1337} & 0.1215 & 0.1293 & 0.1226 & 0.1191 & 0.0083 & \multicolumn{1}{c|}{0.0713} & \textbf{0.1724*} & 28.96\% \\ \bottomrule
\end{tabular}
\caption{Overall performance. The best and second-best are \textbf{bold} and \underline{underlined}. Gain measures the difference between \ours~and the best competitive baseline. `*' indicates statistically significant improvement $(p < 0.01)$ for a paired $t$-test of \ours~compared to the best baseline, as conducted across 5 experiments.}\label{tab:overall_performance}
\end{table*}

\iffalse

\begin{table*}[] \small
\centering
% \setlength{\tabcolsep}{3.4pt}
\renewcommand{\arraystretch}{0.95}
\begin{tabular}{c|c|ccc|cc|cc|cc}
\toprule
\multicolumn{2}{c|}{Model} & \multicolumn{3}{c|}{Traditional CRS} & \multicolumn{2}{c|}{LLM-based CRS} & \multicolumn{2}{c|}{Traditional IR} & \multicolumn{2}{c}{Retrieval-based CRS} \\ \midrule
Dataset & Metric & RevCore & UniCRS & ECR & Zero-shot & ChatCRS & BM25 & DPR & BM25CRS & \aours \\ \midrule
 & R@10 & 0.0268 &  &  & 0.0767 & 0.0763 & 0.0026 & 0.0940 &  &  \\
 & R@50 & 0.0898 &  &  & 0.1129 & 0.1168 & 0.0123 & 0.2206 &  &  \\
 & N@10 & 0.0132 &  &  & 0.0468 & 0.0462 & 0.0014 & 0.0502 &  &  \\
\multirow{-4}{*}{\spearl} & N@50 & 0.0266 &  &  & 0.0560 & 0.0565 & 0.0033 & 0.0777 &  &  \\ \midrule
 & R@10 &  & 0.1330 &  & 0.1436 & 0.1410 &  & 0.1019 &  & 0.3295 \\
 & R@50 &  & 0.2868 &  & 0.2436 & 0.2436 &  & 0.2672 &  & 0.5537 \\
 & N@10 &  & 0.0723 &  & 0.0927 & 0.0806 &  & 0.0512 &  & 0.1720 \\
\multirow{-4}{*}{\sinspired} & N@50 &  & 0.1059 &  & 0.1175 & 0.1071 &  & 0.0872 &  & 0.2217 \\ \midrule
 & R@10 & 0.1739 & 0.1095 &  & 0.1670 & 0.1666 &  & 0.0774 &  &  \\
 & R@50 & 0.3034 & 0.2482 &  & 0.2783 & 0.2824 &  & 0.2138 &  &  \\
 & N@10 & 0.1053 & 0.0613 &  & 0.0937 & 0.0893 &  & 0.0403 &  &  \\
\multirow{-4}{*}{\sredial} & N@50 & 0.1337 & 0.0915 &  & 0.1226 & 0.1191 &  & 0.0713 &  &  \\ \bottomrule
\end{tabular}
\caption{Overall performance. The best and }\label{tab:overall_performance}
\end{table*}

\fi
\section{Experimental Setup}
\label{sec:experimental_setup}
% \section{Experiments Setup}\label{sec:exp_setup}
\subsection{Datasets}
% We validate the performance of R2C on two datasets.
% (i)  \textbf{In-domain}: We utilize Natural Questions (NQ)~\cite{tacl/KwiatkowskiPRCP19/NQ}, which are widely adopted in Open-domain Question Answering (ODQA) tasks. We retrieve 20 candidate passages for each question using DPR~\cite{emnlp/KarpukhinOMLWEC20/DPR, iclr/IzacardG21/FiD-KD}. 
% (ii) \textbf{Out-of-domain}: To evaluate the generalizability of compressed prompts, we use the LongBench~\cite{bai2023longbench}\footnote{\url{https://github.com/THUDM/LongBench}}. We include five types of tasks: single-document QA (SingleDoc), multi-document QA (MultiDoc), summarization (Summ.), few-shot learning (FewShot), and code completion (Code). Note that we only evaluate the English datasets and omit the synthetic tasks to validate the model's ability in real-world scenarios. (See appendix~\ref{sec:app_dataset} for further details.)

% 우리는 \ours의 성능을 평가하기 위해 영화 도메인에 대해 세 가지 벤치마크 데이터셋을 활용한다. \redial\cite{data2018redial}과 \inspired\cite{data2020inspired}는 모두 Amazon Mechanical Turk (AMT) platform에서 crowd-sourcing을 통해 구축된 널리 사용되는 데이터셋이다. \pearl\cite{data2024pearl}은 영화 리뷰를 기반으로 구축된 데이터셋으로 사용자의 persona를 반영하도록 의도된 대화 데이터셋이다. 각 데이터는 그 preference density가 다양하며, 데이터셋의 통계는 Table ~\ref{tab:data_statistics}에서 보여준다.
To evaluate the performance of \ours, we utilize three benchmark datasets in the movie domain. \inspired~\cite{data2020inspired} and \redial~\cite{data2018redial} are widely used datasets built through crowdsourcing on the Amazon Mechanical Turk (AMT) platform. \pearl~\cite{data2024pearl} is a dataset constructed based on movie reviews, designed to reflect the user's persona in conversations. The dataset statistics are summarized in Table~\ref{tab:data_statistics}.

\subsection{Evaluation Protocol}
To evaluate the recommendation performance on the CRS models, we utilize the widely used ranking metrics NDCG@$k$ and Recall@$k$ (with $k=10, 50$). Notably, previous research~\cite{llm/crs2023llmzs} has found that ground-truth items already seen in previous dialogue can lead to shortcuts. Therefore, we exclude these items from the ground-truth set to ensure a more accurate assessment following~\cite{llm/crs2023llmzs,llm/crs2024memocrs, genret/crs2024rta}.
% For Natural Questions, we use Span Exact Match (Span EM) and prompts following \citet{corr/abs-2307-03172/lost-in-the-middle} to evaluate whether the generated text contains the answer. For LongBench, we follow metrics and prompts of each dataset provided by the official benchmark (Refer to appendix~\ref{sec:app_metric} and~\ref{sec:app_prompt}).

\subsection{Baselines}
% We compared \ours~with the following models: \textit{(i) Traditional CRS:} UniCRS~\cite{kg/crs2022unicrs}, RevCore~\cite{pref/crs2021revcore}, and ECR~\cite{pref/crs2024ecr}, which enhance domain-related knowledge through knowledge graphs, with RevCore and ECR considering user sentiments towards extracted entities. While RevCore classifies user sentiment about entities into positive and negative categories, ECR further refines user representation by categorizing sentiment into nine distinct emotions. 
% We evaluate our approach against several models: \textit{(i) Traditional CRS:} UniCRS~\cite{kg/crs2022unicrs}, RevCore~\cite{pref/crs2021revcore}, and ECR~\cite{pref/crs2024ecr}, which incorporate domain-specific knowledge using knowledge graphs. RevCore and ECR additionally analyze user sentiments towards entities; RevCore distinguishes sentiments as positive or negative, while ECR classifies them into nine distinct emotions.
We compare \ours\ with seven baselines.

\begin{itemize}[leftmargin=5mm]
    \item {\textbf{Traditional CRS models}}: \textbf{UniCRS}~\cite{kg/crs2022unicrs}, \textbf{RevCore}~\cite{pref/crs2021revcore}, and \textbf{ECR}~\cite{pref/crs2024ecr} leverage domain-specific knowledge via knowledge graphs. For each entity, RevCore categorizes sentiments into positive or negative, whereas ECR identifies nine distinct emotional responses.

    \vspace{0.5mm}
    \item {\textbf{LLM-based CRS models}}: \textbf{Zero-shot} recommends based solely on dialogue history and internal knowledge of items. \textbf{ChatCRS}~\cite{llm/crs2024chatcrs} enhances the domain knowledge of LLM through a knowledge graph.

    \vspace{0.5mm}
    \item {\textbf{Retrieval-based CRS models}}: \textbf{BM25}~\cite{ir94BM25} ranks items by term relevance from a static index, while \textbf{DPR}~\cite{ir2020dpr} retrieves items based on the similarity with dense vectors of the dialogue context. We use $T_c$ and $T_m$ as the user and item textual representation.

\end{itemize}

\subsection{Implementation Details}
% 본 논문에서는 우리는 retrieval을 위한 encoder로 NV-Embed~\cite{lm2024NVEMBED}를 사용하였으며, parameter efficient fine-tuning 기법인 LoRA~\cite{lm2023LoRA}를 적용하여 학습하였다. 각 데이터셋에 적용한 batch size, negative sample 수, learning rate 등은 Appendix ~\ref{}에서 상술한다.
% In this paper, ChatGPT(\texttt{gpt-4o-mini}) in the recommendation-tailored augmentation stage to extract user preferences. NV-Embed~\cite{lm2024NVEMBED} was utilized as the encoder for retrieval, and we applied LoRA~\cite{nlp@2021LoRA}, a parameter-efficient fine-tuning technique for training. $\alpha$는 0.5로 설정되었고, $\beta$는 \pearl, \inspired, \redial에 대해 각각 0.3, 0.2, 0.1로 설정되었다. Detailed configurations, including batch size, number of negative samples, and learning rate for each dataset, are provided in Appendix~\ref{}. 
We utilize \texttt{gpt-4o-mini} for contrasting preference expansion. We used \texttt{gpt-4o-mini-2024-07-18} in all of our experiments including baselines. We initialize the model parameters with NV-Embed~\cite{lm2024NVEMBED} (NV-Emb.) where $d = 4096$, and we applied LoRA~\cite{llm@LoRA}, a parameter-efficient fine-tuning technique for training. The parameter $\alpha$ was set to 0.5, and $\beta$ was set to 0.3, 0.1, and 0.2 for the \inspired, \redial, and \pearl~dataset, respectively. We set the batch size to 8 for \pearl, and 10 for \inspired~and \redial. For negative samples, we used 24 for \pearl\ and 16 for \inspired~and \redial. We use Adam optimizer~\cite{imple2015Adam} with a learning rate of 5e-5 for \pearl~and \redial, and 1e-4 for \inspired. We adopt early stopping based on NDCG@10, with a patience of 3 for \pearl, and 5 for \inspired~and \redial. The temperature parameter $\tau$ is set to 0.05.
We set the maximum sequence length to 512 for items and conversations and 256 for likes and dislikes. The warm-up steps are set to 10\% of one epoch.
We set $k=3$ for the review summarization, using 3 reviews per item. The prompts used in Contrasting Preference Expansion and examples are provided in Appendix~\ref{sec:appendix_prompt} and detailed configurations for the baselines are provided in Appendix~\ref{sec:appendix_implementation_baseline}.
% For a fair comparison, all LLM-based models were evaluated using \texttt{gpt-4o-mini}.

% batch size는 \pearl은 8, \inspired랑 \redial는 10
% negative sample은 \pearl은 24, \inspired랑 \redial은 16
% learning rate는 \pearl, \redial은 5e-5, \inspired은 1e-4
% early stop with ndcg@10, patience 3 for \pearl, 5 for \inspired and \redial
% temp set to 0.05
% max length는 512 for item, conv, 256 for like and dislike
% warmup step은 한 에폭의 0.1
% review summary를 위한 review는 3개를 사용한다
\begin{figure}[t]
\centering
\includegraphics[width=1\linewidth]{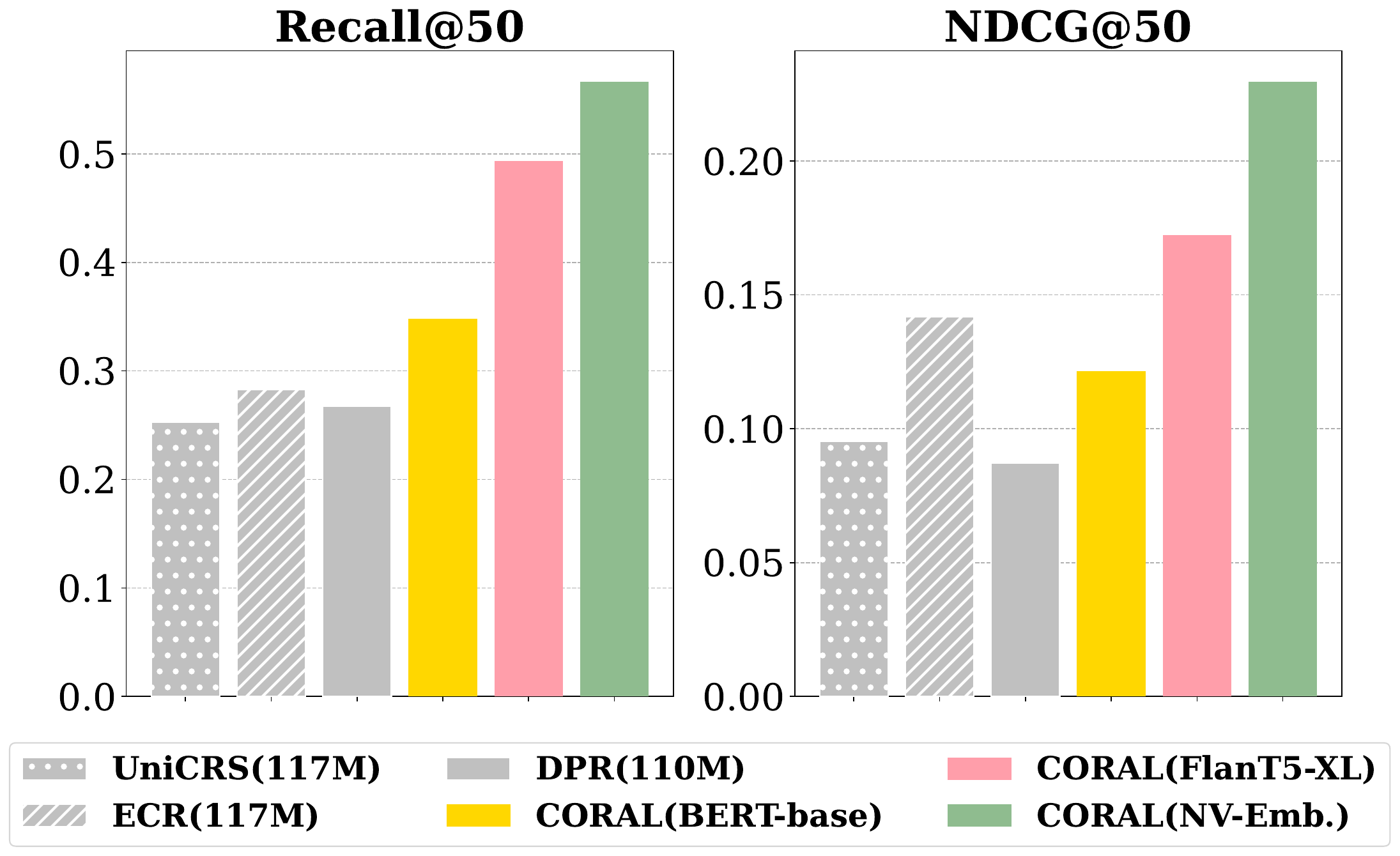} 
\caption{Performance according to various model sizes in \inspired.}
\label{fig:slm}
\end{figure}

\section{Results and Analysis}
\label{sec:results}

\subsection{Overall Performance}
Table~\ref{tab:overall_performance} compares \ours~against seven baselines across three datasets.
% 전반적인 성능
\ours~achieves state-of-the-art performance, improving Recall@50 and NDCG@50 over the best competitive baseline by an average of 43.87\% and 48.59\%, respectively. This demonstrates that it consistently captures and models user preferences across datasets with diverse characteristics.
% 가장 가까운 baseline과의 비교
Notably, \ours~outperforms ECR~\cite{pref/crs2024ecr}, which enhances a single user representation with sentiment. This indicates that explicitly learning the relationships between preferences and users/items is more effective for user preference modeling.

\begin{table*}[]
\centering
\resizebox{\textwidth}{!}{ % 테이블 크기를 줄임
\begin{tabular}{cc|rcrc|cccc|cccc}
\toprule
\multicolumn{2}{c|}{Dataset} &
  \multicolumn{4}{c|}{\pearl} &
  \multicolumn{4}{c|}{\inspired} &
  \multicolumn{4}{c}{\redial} \\ \midrule
\multicolumn{1}{c|}{Retriever} &
  $L, D$ &
  \multicolumn{1}{l}{R@10} &
  R@50 &
  \multicolumn{1}{l}{N@10} &
  N@50 &
  \multicolumn{1}{l}{R@10} &
  R@50 &
  \multicolumn{1}{l}{N@10} &
  N@50 &
  \multicolumn{1}{l}{R@10} &
  R@50 &
  \multicolumn{1}{l}{N@10} &
  N@50 \\ \midrule
\multicolumn{1}{c|}{\multirow{2}{*}{BM25}} &
  w/o &
  0.0053 &
  0.0269 &
  0.0031 &
  0.0076 &
  0.0390 &
  0.1486 &
  0.0216 &
  0.0444 &
  0.0091 &
  0.0361 &
  0.0044 &
  0.0104 \\
\multicolumn{1}{c|}{} &
  w/ &
  \textbf{0.0079} &
  \textbf{0.0340} &
  \textbf{0.0045} &
  \textbf{0.0099} &
  \textbf{0.0448} &
  \textbf{0.1714} &
  \textbf{0.0297} &
  \textbf{0.0580} &
  \textbf{0.0235} &
  \textbf{0.0704} &
  \textbf{0.0113} &
  \textbf{0.0219} \\ \midrule
\multicolumn{1}{c|}{\multirow{2}{*}{BERT}} &
  w/o &
  \textbf{0.0040} &
  0.0203 &
  \textbf{0.0015} &
  0.0049 &
  0.0057 &
  0.0410 &
  0.0017 &
  0.0100 &
  0.0069 &
  0.0270 &
  0.0039 &
  0.0084 \\
\multicolumn{1}{c|}{} &
  w/ &
  0.0031 &
  \textbf{0.0225} &
  \textbf{0.0015} &
  \textbf{0.0056} &
  \textbf{0.0086} &
  \textbf{0.0648} &
  \textbf{0.0032} &
  \textbf{0.0152} &
  \textbf{0.0087} &
  \textbf{0.0302} &
  \textbf{0.0048} &
  \textbf{0.0096} \\ \midrule
\multicolumn{1}{c|}{\multirow{2}{*}{NV-Emb.}} &
  w/o &
  0.0454 &
  0.1323 &
  0.0210 &
  0.0397 &
  0.1648 &
  0.3943 &
  0.0856 &
  0.1374 &
  0.0767 &
  0.1885 &
  0.0368 &
  0.0613 \\
\multicolumn{1}{c|}{} &
  w/ &
  \textbf{0.0569} &
  \textbf{0.1508} &
  \textbf{0.0286} &
  \textbf{0.0489} &
  \textbf{0.2276} &
  \textbf{0.4048} &
  \textbf{0.1140} &
  \textbf{0.1538} &
  \textbf{0.0872} &
  \textbf{0.2190} &
  \textbf{0.0408} &
  \textbf{0.0711} \\ \bottomrule
\end{tabular}
}
\caption{The zero-shot performance of various language models depending on the presence or absence of the user's potential preference. $L$ and $D$ mean $T^{\mathrm{aug}}_l$ and $T^{\mathrm{aug}}_d$, respectively.}
\label{tab:lm_zero_shot}
\end{table*}

% \usepackage{multirow}
% \usepackage{multirow}
\begin{table}[] \small
\centering
\begin{tabular}{c|cccc}
\toprule
Dataset & \multicolumn{4}{c}{\sinspired} \\ \midrule
Variants & R@10 & \multicolumn{1}{c|}{R@50} & N@10 & N@50 \\ \midrule
\sours & \textbf{0.3481} & \multicolumn{1}{c|}{0.5667} & \textbf{0.1827} & \textbf{0.2297} \\ \midrule
w/o $L, D$ & 0.3248 & \multicolumn{1}{c|}{\textbf{0.5767}} & 0.1668 & 0.2226 \\
w/o $R$ & 0.3167 & \multicolumn{1}{c|}{0.5348} & 0.1710 & 0.2193 \\
w/o $L, D, R$ & 0.2948 & \multicolumn{1}{c|}{0.5633} & 0.1595 & 0.2196 \\
w/o $Aug.$ & 0.3016 & \multicolumn{1}{c|}{0.5379} & 0.1663 & 0.2202 \\ \midrule
w/o $Neg.$ & 0.2974 & \multicolumn{1}{c|}{0.5692} & 0.1520 & 0.2115 \\
w/o $PL$ & 0.1847 & \multicolumn{1}{c|}{0.3616} & 0.1107 & 0.1491 \\ \bottomrule
\end{tabular}
\caption{Ablation study of \ours~in \inspired. The best scores are in \textbf{bold}. $L$, $D$ and $R$ denote $T_{\mathit{l}}^{\mathrm{aug}}$, $T_{\mathit{d}}^{\mathrm{aug}}$, and $T_{\mathit{r}}^{\mathrm{key}}$, respectively. Also, \emph{Aug.}, \emph{Neg.}, and \emph{PL} mean potential preference augmentation, hard negative sampling, and preference-aware learning, respectively.
}\label{tab:Ablation study}
\end{table}

\iffalse

\begin{table*}[] \small
\centering
\begin{tabular}{cc|cccc|cccc}
\toprule
\multicolumn{2}{c|}{Dataset}                                                      & \multicolumn{4}{c|}{\pearl}        & \multicolumn{4}{c}{\inspired}      \\ \midrule
\multicolumn{2}{c|}{Metric}                                                       & R@10   & R@50   & N@10   & N@50   & R@10   & R@50   & N@10   & N@50   \\ \midrule
\multicolumn{2}{c|}{Ours}                                                         &        &        &        &        & 0.3481 & 0.5667 & 0.1827 & 0.2297 \\ \midrule
\multicolumn{1}{c|}{\multirow{3}{*}{Input information}} & w/o \textit{like, dislike} &        &        &        &        & 0.3248 & 0.5767 & 0.1668 & 0.2226 \\
\multicolumn{1}{c|}{}                                & w/o \textit{reviews}        & 6      &        &        &        & 0.3167 & 0.5348 & 0.1710 & 0.2193 \\
\multicolumn{1}{c|}{}                                & w/o \textit{like, dislike, review     }& 5      &        &        &        & 0.2948 & 0.5633 & 0.1595 & 0.2196 \\ \midrule
\multicolumn{1}{c|}{\multirow{2}{*}{Model ablation}} & w/o \textit{ranking}       & 2      &        &        &        &        &        &        &        \\
\multicolumn{1}{c|}{}                                & w/o neg. sampling & 0.1605 & 0.3122 & 0.0962 & 0.1354 & 0.2974 & 0.5692 & 0.1520 & 0.2115 \\ \bottomrule
\end{tabular}
\caption{Overall performance}\label{tab:Abalation study}
\end{table*}

\fi

% sentiment 쓰는 baseline들
RevCore~\cite{pref/crs2021revcore} and ECR~\cite{pref/crs2024ecr}, which utilize user sentiment, outperform UniCRS~\cite{kg/crs2022unicrs} in \inspired~and \redial~but not in \pearl. One possible reason is relatively longer dialogues in \pearl, which demand a more contextual ability to capture conversational context. 
RevCore and ECR focus on sentiment extraction at the entity and utterance levels, respectively, making it challenging to capture sentiment considering the full context.
In contrast, \ours\ identifies user sentiment at the conversation level, achieving a more comprehensive understanding of user preferences throughout the dialogue.
% However, \ours~identifies user sentiment at the conversation-level utilizing the reasoning capabilities of LLM, thereby achieving a more comprehensive understanding of user preference within the dialogue.

\subsection{In-depth Analysis}
\subsubsection{Performance by Model Size}
Figure~\ref{fig:slm} presents the performance of \ours~with different backbone model sizes.
We have two observations as follows.
(i) The performance increases as the model size increases, \ie, BERT-base (110M) $\rightarrow$ FlanT5-XL (1.5B) $\rightarrow$ NV-Embed (7B). This is because larger models are better at capturing complex semantic relationships in text. Additionally, \ours~leverages a dense retriever structure to fully exploit the capabilities of PLMs.
% 이것은 큰 모델일수록 텍스트의 복잡한 의미적 관계를 더 잘 포착할 수 있고, \ours가 dense retriever 구조를 사용하여 take full advantage of PLM이기 때문이다.
(ii) \ours~significantly improves performance even with a relatively small model. Both DPR and \ours~(BERT-base) use the same backbone. \ours~(BERT-base) shows a 39\% performance gain in NDCG@50 and achieves comparable performance to similarly sized baselines, such as UniCRS and ECR.
% When comparing DPR and \ours(BERT-base), both use the same backbone, \ours~shows a 39\% performance gain in NDCG@50.
These reveal that \ours~is a highly scalable, universal, and effective framework that can be applied to any model size.
% (i) LLM 모델 크기가 커질 수록 성능이 올라간다. Model size가 BERT-base (110M), FlanT5-XL (1.5B), NV-Embed (7B)으로 커질수록 성능이 올라간다.
% (ii) 같은 backbone을 공유하는 DPR과 비교했을 때, \ours가 NDCG@50에서 39\%의 gain을 보여준다.
% 이러한 결과는 \ours가 scalability가 뛰어나며, 모델 크기에 상관없이 적용될 수 있는 범용적이고 effective한 framework임을 드러낸다.
% 또한 비슷한 모델 크기에서 \ours~(BERT)가 DPR~(BERT)와 ECR과 비슷하거나 더 높은 성능을 보인다.

\subsubsection{Zero-shot Performance}
% Table~\ref{tab:lm_zero_shot} evaluates the zero-shot performance of \ours~on a sparse retriever and two dense retrievers of different sizes.
% 이때 ambivalent preference의 textual representation 대신 $T_m$과 ${T_r}^{\mathrm{key}}$를 concat하여 사용하였다.
% \ours~significantly improves performance independent of the backbone, achieving a 37\% performance gain on average.
% 이는 두 가지를 의미한다.
% (i) 우리가 추출한 user preference가 다양한 retriever 구조와 size에서도 effective하다.
% (ii) Similarity scoring이 학습 없이도 conv와 like, dislike 그리고 item 간 관계를 효과적으로 모델링한다.
To investigate the effectiveness of user-side expansion, we compare the zero-shot performance of various models with and without leveraging the user's potential preferences.
% To investigate the effectiveness of user-side expansion, we compare the zero-shot performance of \ours~ w
We evaluate zero-shot performance on a sparse retriever and two dense retrievers with varying sizes, as shown in Table~\ref{tab:lm_zero_shot}.
For `\emph{w/o $L$, $D$}', we only utilize $T_c$ for user-side textual representation without $T_{\mathit{l}}^{\mathrm{aug}}$, and $T_{\mathit{d}}^{\mathrm{aug}}$.
`\emph{w/ $L$, $D$}' utilizes $T_c, T_{\mathit{l}}^{\mathrm{aug}}, T_{\mathit{d}}^{\mathrm{aug}}$ and the score is computed as Equation~\eqref{eq:similarity_scoring}.
For the item-side textual representation, we concatenated $T_m$ with ${T_r}^{\mathrm{key}}$.
\ours~significantly improves performance across different backbones, demonstrating a 37\% average gain in the zero-shot setting.
These results confirm that contrasting preference expansion effectively improves recommendation performance by inferring potential user preferences. Appendix~\ref{sec:appendix_zero_shot_ablation} provides more detailed ablation for user-side expansion.

% Table~\ref{tab:lm_zero_shot}에서, \ours~significantly improves performance independent of the backbone, achieving a 37\% performance gain on average in zero-shot setting. 이때 ambivalent preference의 textual representation 대신 $T_m$과 ${T_r}^{\mathrm{key}}$를 concat하여 사용하였다. 
% We evaluate the zero-shot performance of \ours~on a sparse retriever and two dense retrievers of different sizes. 실험 결과는 우리가 증강한 potential user preference가 추천 성능을 개선하는 recommendation-tailored augmentation approach임을 입증한다. like/dislike의 다양한 ablation setting에서의 결과는 Table~\ref{tab:zs_ablation}에서 확인할 수 있다.

\begin{figure*}[t]
\centering
\includegraphics[width=1.0\linewidth]{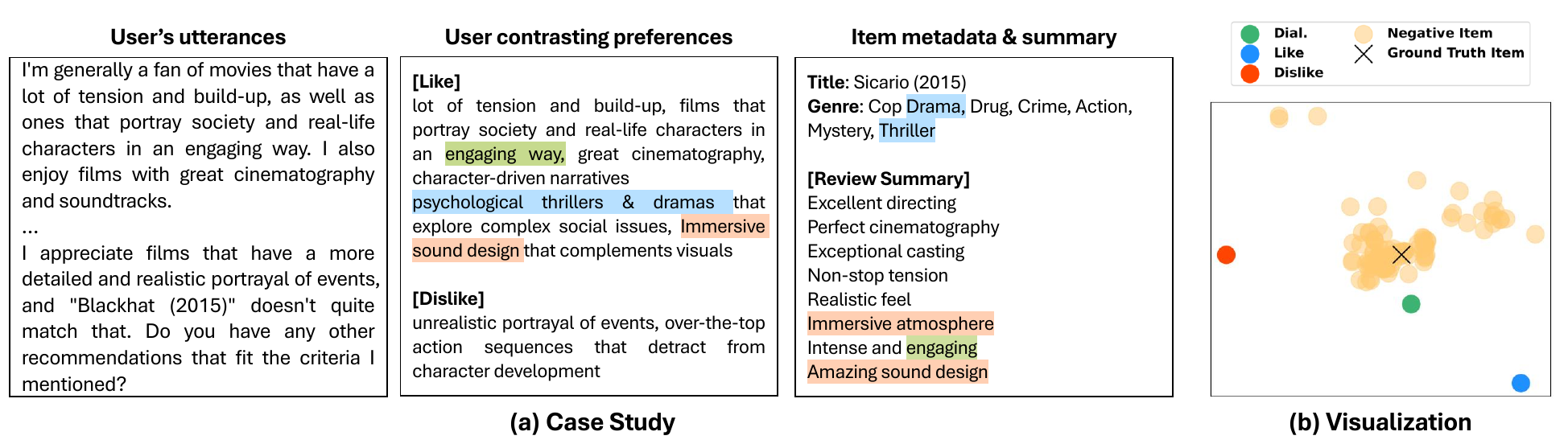} 
\caption{Case study and visualization of \ours\ in \pearl~dataset.}
\label{fig:case_study}
\end{figure*}

\subsubsection{Ablation Study}
To understand the impact of each component of \ours~on performance, we conducted an ablation study on \inspired, as illustrated in Table~\ref{tab:Ablation study}.
First, we validate the effectiveness of contrasting preference expansion. `\emph{w/o $L,D$}' means that $T_{\mathit{l}}^{\mathrm{aug}}$ and $T_{\mathit{d}}^{\mathrm{aug}}$ were not used in both train and inference.
We can see that all three expansions (\ie, $L$, $D$, and $R$) contribute to performance, especially at a high ranking. It indicates that the expansions allow us to distinguish subtle preferences.
We then investigate the effect of augmenting user's potential preferences.
For `\emph{w/o Aug.}', we use $T_{\mathit{l}}^{\mathrm{ext}}$, $T_{\mathit{d}}^{\mathrm{ext}}$ as the user's positive and negative preferences with only superficial preference extraction.
We find that augmenting the potential preference significantly improves performance, yielding up to a 15.41\% increase in Recall@10. It highlights the importance of underlying preference within the dialogue.

Lastly, we explore the effects of our proposed preference-aware learning.
`\emph{w/o Neg.}' is a variant that uses in-batch negative instead of hard negative, and `\emph{w/o PL}' is a variant without preference modeling and hard negative, which utilizes single user representation by concatenating $T_c, T_{\mathit{l}}^{\mathrm{aug}}, T_{\mathit{d}}^{\mathrm{aug}}$.
Compared to \ours, removing negative sampling or preference modeling leads to a significant drop in performance. Hence, preference-aware learning effectively learns the relationship between conversation, preferences, and items. Refer to Appendix~\ref{sec:appendix_slm_ablation} for the results of the ablation study on BERT, and Appendix~\ref{sec:appendix_other_llm} for the results using different LLMs in Contrasting Preference Expansion.

\subsection{Case Study and Visualization} \label{sec:case_study}
Figure~\ref{fig:case_study}(a) shows a case study of the user's dialogue, the augmented preferences, and the ground truth item from \pearl.
The highlighted phrases in the same color represent phrases that belong to the same concept and are related to the ground-truth item. 
In particular, the blue, green, and orange preferences mean newly augmented through contrasting preference expansion.

% User side expansion
Firstly, \ours~effectively inferred the user's preference for \emph{Thriller} and \emph{Drama} genres based on the dialogue.
This demonstrates that leveraging the augmentation effectively makes it possible to predict user preferences that are difficult to infer solely from the given dialogue history.
% Item side expansion
Also, it can be observed that the user's preferences align with the item’s review summary (\eg, \textit{immersive sound design} in user preference, \textit{immersive atmosphere} in item review summary). This indicates that the review summary successfully bridges the gap between the preferences expressed in the user conversation and the item.
% Explainability
Additionally, \ours's contrasting preference expansion serves as a rationale for the recommendation results, thereby providing explainability.

We then visualized the corresponding example in the embedding space using t-SNE in Figure~\ref{fig:case_study}(b), to illustrate how \ours~utilizes preference.
Green, blue, and red dots are the embedding vectors of $h_c$, $h_l$, $h_d$ for the dialogue shown in Figure~\ref{fig:case_study}(a).
The X marks indicate the ground truth item and the orange dots represent negative items that are not the ground truth.
The example shows incorrect items, such as those close to the dialogue but positioned toward the ``dislike'' items, that may be selected when only the conversation is used.
This is because user conversations contain contrasting preferences.
% As shown in the example, sub-optimal items may be selected when only the conversation is used.
However, since \ours~explicitly differentiates like/dislike preferences and models the relationship between the user and the item, it successfully recommended the correct item by distinguishing subtle differences.
% 이런 식으로 다시 작성해볼게요 (내일 오후까진 완성해볼게요 감사합니다)

% 1. CRS 설명, 두 가지 Task 설명
% 2. Traditional CRS 모델 소개 및 한계점 / Preference-aware CRS 모델 언급
% 3. Existing Preference-aware CRS 모델 소개 및 한계점
% 4. 두 가지 Question 던지기
%    4-1. Q1: 사용자가 좋아하고 싫어하는걸 어떻게 알거고, 이걸 어떻게 표현할거임? (how to represent)
%    4-2. Q2: 사용자가 좋아하고 싫어하는 걸 어떻게 item과 align해서 추천할거임? (how to align)
% 5. 제안 방법 소개
%    5-1. A1: 추출하고 증강하고 text로 표현 (semantic, context 키워드 강점)
%    5-2. A2: representation learnining (user <-> positive preference and negative preference <-> item)
%         - retrieval-based CRSs의 이점 언급
% 6. Contribution

% 구성 계획
% Traditional CRS 소개하며, Preference 고려하는 것의 한계와 중요성 언급 후, 아래 카테고리로 전개
% 1. Preference-aware CRS.
% 2. Retrieval-based CRS
% ours와의 차별점을 위 카테고리 각각에 짧게 넣을지, 마지막에 종합적으로 넣을지 고려 중에 있음

\section{Related Work}
\label{sec:related_work}
% 1-1) KG를 사용하는 모델
Traditional conversational recommender systems (CRSs)~\cite{survey/crs2021CRSsurvey} increasingly harness external information, such as knowledge graphs~\cite{kg/crs2019kbrd, kg/crs2020kgsf, kg/crs2024kascrs, kg/crs2022unicrs, kg/crs2023c2crs} and metadata~\cite{meta/crs2022mese}, to enhance domain knowledge.
% 1-2) LLM과 KG를 사용하여 context knowledge를 보강하려는 연구
With the substantial impact of large language models (LLMs) demonstrating exceptional world knowledge, recent studies~\cite{llm/crs2023llmzs, rs2023chatgptgoodrec, llm/crs2024chatcrs, llm/crs2024feedbackcrs} have focused on utilizing LLMs as standalone recommenders. In particular, several approaches~\cite{llm/crs2021kgensam, llm/crs2024chatcrs} have been proposed to integrate the strong contextual understanding of LLMs with knowledge graphs to address gaps in domain-specific knowledge to improve the system's overall performance.

% 2) 기존 CRS models의 한계점
% However, these studies often overlook the fact that users in dialogue may have varied emotions and attitudes toward each entity mentioned. This oversight is significant because they fail to fully capture the complexity of the conversation, thereby harming user experience.
However, these studies commonly neglect the diversity of users' emotions and attitudes toward entities in dialogues, which undermines the conversation's complexity and degrades user experience.

\vspace{0.5mm}
\noindent
\textbf{Preference-aware CRSs.}
% 기존 preference-aware CRS models 소개
Several studies have focused on improving user preferences at the entity level by considering user emotions using knowledge graphs~\cite{pref/crs2021revcore, pref/crs2024ecr}. RevCore~\cite{pref/crs2021revcore} classifies entities from conversations as positive or negative and retrieves emotion-related reviews to enhance user expressiveness. ECR~\cite{pref/crs2024ecr} leverages LLMs to categorize entities into nine specific emotional categories, segmenting the user preferences. MemoCRS~\cite{llm/crs2024memocrs} employs a memory-enhanced approach to ensure preference continuity by tracking sequential user preferences. Although these studies consider user preferences at various levels of granularity and context, they still overlook the existence of contrasting preferences. 
% Failing to account for these differing preferences can lead to inaccurate reflections of user intent during item recommendation, ultimately resulting in suboptimal recommendation outcomes.

% 위의 연구들이 사용자의 선호를 다양한 세분성 수준과 context를 고려하고 있지만, 여전히 각각의 연구는 서로 상충되는 선호 정보—즉, positive와 negative—의 존재를 간과하고 있다. 이런 선호 정보를 서로 다르게 고려하지 않는 것은 아이템 추천 시 사용자의 의도를 정확하게 반영하지 못해 결과적으로 suboptimal한 추천 결과를 도출할 수 있다.

\vspace{0.5mm}
\noindent
\textbf{Retrieval-based CRSs.}
Recent work~\cite{ret/rs2024blair, data2024AlignLM2IR, ret/crs2023CRSRetbaseline, ret/crs2024nls} reformulated recommender systems as item retrieval tasks, fully utilizing the semantic understanding and matching capabilities of language models. In light of this, a few studies~\cite{ret/crs2023CRSRetbaseline, ret/crs2024nls} leveraging retrievers have been introduced to enhance CRS tasks. Specifically, they treat the conversation as a query and items as documents and utilize text-matching techniques such as BM25~\cite{bm252021pyserini}, offering high generalizability and scalability. 

\section{Conclusion}
\label{sec:conclusion}
% 우리 연구 요약
% 각 컴포넌트 각각 설명 후 contribution과 실험 결과
% In this paper, 우리는 retriever 아키텍쳐에서 ambivalent preferences 모델링을 탐구하였다. We propose a novel framework, \ours, which extracts users' like/dislike preferences from their dialogue history and explicitly incorporates this information into the training process, enabling the retriever to consider ambivalent preferences when searching for items.
% 특히, preference expansion phase에서는 reasoning power of LLMs을 활용하여 대화 내 사용자의 current preferences를 정확하게 추출 및 분류할 뿐만 아니라, 미래에 사용자가 좋아할 법한 preference를 추론하여 확장함으로써 recommendation에 도움이되는 추천 맞춤형 증강을 사용할 수 있으며, ablation study의 
% Section~\ref{tab:Ablation study}에서 증강된 선호가 유효함을 실험적으로 증명하였다.
% Preference-aware learning에서 ambivalent한 선호를 서로 분리하여 학습하고 사용자와 ambivalent한 선호, 그리고 아이템 사이의 관게를 직접적으로 학습하여 직접 item scoring에 관여할 수 있도록 한다. 실험 결과, \ours는 전체 베이스라인을 outperform하며 다양한 크기의 backbone language model에 대해서도 강건한 성능을 보인다.
% 우리는 공식 코드와 세 개의 벤치마크 대화 데이터셋에 대해 생성한 사용자의 like/dislike preferences를 공개할 예정이다.

% In this paper, we proposed \ours, a novel CRS framework that enhances personalized recommendations by modeling ambivalent preferences from users' dialogue histories. Firstly, we augmented users' current and potential future preferences with the reasoning capabilities of LLMs. With the extracted potential preferences, we learn the relationship between the conversation and items on top of retriever architectures.
% Experimental results confirm the effectiveness of our approach, showing that \ours\ outperforms all baselines across various language models.

In this paper, we proposed a new CRS model \ours~that distinguishes users' ambiguous preferences, implying conflicting intentions inherent in their likes and dislikes. To support the recommendation task, \ours\ expands users' preferences and leverages the reasoning capabilities of LLMs to learn the relationships between conversations, preferences, and items. Extensive experiments are conducted to evaluate the effectiveness of our preference expansion and learning strategy, confirming that our approach surpasses all baseline models in enhancing recommendation performance (Table~\ref{tab:overall_performance}). \ours\ can also be robustly and universally applied across various language models (Figure~\ref{fig:slm}), and operates effectively in a zero-shot setting, demonstrating the reliability of our augmented user potential preferences (Tables~\ref{tab:lm_zero_shot}~and~\ref{tab:zs_ablation}).

% In this paper, we proposed \ours, a novel CRS framework that models ambivalent preferences by extracting like/dislike preferences from users' dialogue histories and incorporating them into the training of retriever architectures. Our framework enhances personalized recommendations by expanding users' current and potential future preferences with the reasoning capabilities of LLMs. Experimental results confirm the effectiveness of our approach, showing that \ours\ outperforms all baselines across various language models.

% We will release the code and user preference data from three benchmark dialogue datasets.
\section*{Limitation}
\label{sec:limitation}
% 본 연구의 한계는 두 가지로 구분할 수 있다: (i) 학습 및 추론 효율성, (ii) LLM 의존성.
% 먼저, 본 연구에서는 NV-Embed (7B)를 parameter-efficient training technique인 LoRA~\cite{nlp@2021LoRA}를 적용하여 10M만 학습하였으며, negative sampling을 통해 학습 효율성을 높였음에도 불구하고, 여전히 large models에서는 학습 시간이 길어지고 비용이 많이 소모된다.
% 또한, ambivalent preference expansion 단계에서 대화로부터 사용자의 선호를 추출하는 과정은 LLM의 reasoning power에 크게 의존한다는 점에서 challenging하다. 이는 추출된 선호의 정확도와 potential preferences의 퀄리티에 중요한 영향을 미치며, LLM의 성능 한계, 크기, bias, 그리고 내포된 지식의 범위에 따라 그 퀄리티가 좌우될 수 있다. 따라서 LLM의 이러한 특성이 학습 결과에 미치는 영향을 면밀히 조사할 필요성이 있다.
The limitations of this study can be categorized into aspects:  (i) training and inference inefficiency and (ii) reliance on large language models (LLMs).

Firstly, in the contrasting preference expansion, extracting user preferences from dialogue heavily depends on the reasoning power of LLMs, which presents a challenge. The accuracy of extracted preferences and the quality of potential preferences are significantly influenced by the performance, size, bias, and knowledge scope of the LLM. Consequently, it is essential to closely examine the impact of these LLM characteristics on the learning outcomes.
Secondly, in this work, we trained NV-Embed (7B) with only 10M parameters using the parameter-efficient training technique LoRA~\cite{llm@LoRA} and enhanced training efficiency through negative sampling. Despite these efforts, training time and computational costs remain high when using LLMs.
% Future work 제시 필요
% 생성 태스크 안함 언급.

\section*{Acknowledgments} 
This work was partly supported by the Institute of Information \& Communications Technology Planning \& Evaluation (IITP)-ICT Creative Consilience Program grant funded by the Korea government (MSIT) (No. RS-2019-II190421, RS-2022-II220680, IITP-2025-RS-2020-II201821, IITP-2025-RS-2024-00360227).

% \bibliographystyle{acl_natbib}
\bibliography{references}

\begin{table*}[]
\centering
\begin{tabular}{@{}c|c|cccc|c@{}}
\toprule
Dataset  & Input info. & R@10   & R@50   & N@10   & N@50   & Avg. Gain(\%)                          \\ \midrule
\pearl    & $C$           & 0.0476 & 0.1230  & 0.0229 & 0.0395 & -                                     \\
         & $C$, $L$        & 0.0560  & 0.1349 & 0.0303 & 0.0474 & \textbf{19.91\%}\\
         & $C$, $D$        & 0.0481 & 0.1349 & 0.0224 & 0.0415 &  \textbf{3.40\%}  \\
         & $C$, $L$, $D$     & 0.0573 & 0.1481 & 0.0311 & 0.0504 &  \textbf{26.05\%} \\ \midrule
\inspired & $C$           & 0.1837 & 0.4133 & 0.1038 & 0.1545 & -                                     \\
         & $C$, $L$        & 0.2103 & 0.3949 & 0.1133 & 0.1534 &  \textbf{4.62\%}  \\
         & $C$, $D$        & 0.2000 & 0.4154 & 0.1064 & 0.1527 &  \textbf{2.68\%}  \\
         & $C$, $L$, $D$     & 0.2205 & 0.4205 & 0.1166 & 0.1597 &  \textbf{9.37\%}  \\ \midrule
\redial   & $C$,          & 0.0887 & 0.2067 & 0.0383 & 0.0640  & -                                     \\
         & $C$, $L$        & 0.0954 & 0.2325 & 0.0415 & 0.0710  &  \textbf{9.83\%}  \\
         & $C$, $D$        & 0.0910  & 0.2129 & 0.0400   & 0.0665 &  \textbf{3.48\%}  \\
         & $C$, $L$, $D$     & 0.0986 & 0.2431 & 0.0427 & 0.0735 &  \textbf{13.78\%} \\
\bottomrule
\end{tabular}
\caption{Zero-shot performance for preference input variant. $C$, $L$, and $D$ denote $T_c$, $T_{\mathit{l}}^{\mathrm{aug}}$, and $T_{\mathit{d}}^{\mathrm{aug}}$, respectively.}
\label{tab:zs_ablation}
\end{table*}

% \usepackage{booktabs}
% \usepackage{multirow}
% \usepackage{graphicx}

\begin{table*}[]
\centering
\begin{tabular}{l|rrrr|rrrr}
\toprule
Dataset     & \multicolumn{4}{c|}{\inspired}                  & \multicolumn{4}{c}{\redial}                 \\ \midrule
Variants &
  \multicolumn{1}{l}{R@10} &
  \multicolumn{1}{l}{R@50} &
  \multicolumn{1}{l}{N@10} &
  \multicolumn{1}{l|}{N@50} &
  \multicolumn{1}{l}{R@10} &
  \multicolumn{1}{l}{R@50} &
  \multicolumn{1}{l}{N@10} &
  \multicolumn{1}{l}{N@50} \\ \midrule
\ours &
  \textbf{0.1219} &
  \textbf{0.3714} &
  0.0625 &
  \textbf{0.1173} &
  \textbf{0.0856} &
  0.2255 &
  \textbf{0.0446} &
  \textbf{0.0765} \\ \midrule
w/o \textit{L, D}    & 0.0962 & 0.3429 & 0.0429          & 0.0968 & 0.0775 & \textbf{0.2287} & 0.0407 & 0.0754 \\
w/o \textit{R}       & 0.1133 & 0.2695 & \textbf{0.0752} & 0.1103 & 0.0749 & 0.2052          & 0.0405 & 0.0702 \\
w/o \textit{L, D, R} & 0.0876 & 0.3010 & 0.0450          & 0.0967 & 0.0764 & 0.2231          & 0.0403 & 0.0738 \\ \midrule
w/o \textit{Neg.}    & 0.0867 & 0.2629 & 0.0371          & 0.0757 & 0.0725 & 0.2151          & 0.0375 & 0.0697 \\
w/o \textit{PL}      & 0.1076 & 0.2467 & 0.0516          & 0.0832 & 0.0774 & 0.2138          & 0.0403 & 0.0713 \\ \bottomrule
\end{tabular}
\caption{Ablation study of \ours~in \inspired~and \redial~on BERT. The best scores are in \textbf{bold}. $L$, $D$ and $R$ denote $T_{\mathit{l}}^{\mathrm{aug}}$, $T_{\mathit{d}}^{\mathrm{aug}}$, and $T_{\mathit{r}}^{\mathrm{key}}$, respectively. Also, \emph{Neg.} and \emph{PL} mean potential hard negative sampling and preference-aware learning, respectively.}
\label{tab:slm_ablation}
\end{table*}

\begin{table*}[]
\centering
\begin{tabular}{c|cccc}
\toprule
Dataset & \multicolumn{4}{c}{\inspired} \\ \midrule
Model & R@10 & R@50 & N@10 & N@50 \\ \midrule
UniCRS & 0.1113 & 0.2528 & 0.0642 & 0.0952 \\
ECR & 0.1711 & 0.2826 & 0.1077 & 0.1417 \\ \midrule
$\text{{C{\small ORAL}}}_{\text{ w/o}~L, D, R}$ & 0.2948 & 0.5633 & 0.1595 & 0.2196 \\
$\text{{C{\small ORAL}}}_{\text{ Mistral}}$ & 0.3162 & 0.5410 & 0.1809 & 0.2326 \\
$\text{{C{\small ORAL}}}_{\text{ gpt-4o-mini}}$ & 0.3417 & 0.5632 & 0.1772 & 0.2255 \\ \bottomrule
\end{tabular}
\caption{Performance depending on the LLM utilized for Contrasting Preference Expansion. $L$, $D$ and $R$ denote $T_{\mathit{l}}^{\mathrm{aug}}$, $T_{\mathit{d}}^{\mathrm{aug}}$, and $T_{\mathit{r}}^{\mathrm{key}}$, respectively.}
\label{tab:appendix_other_llm}
\vspace{-3mm}
\end{table*}

% \usepackage{multirow}
\begin{table}[] \small
\begin{tabular}{cc|cc|c}
\toprule
\multicolumn{2}{c|}{Model}                              & \multicolumn{2}{c|}{LLM zero-shot} & \multirow{2}{*}{\ours} \\ 
\multicolumn{1}{c|}{Dataset}                   & Metric & \texttt{4o-mini}    & \texttt{4o}            &                       \\ \midrule
\multicolumn{1}{c|}{\multirow{4}{*}{\spearl}}    & R@10   & 0.0767         & 0.1398            & \textbf{0.1851}       \\
\multicolumn{1}{c|}{}                          & R@50   & 0.1129         & 0.1869            & \textbf{0.3619}       \\
\multicolumn{1}{c|}{}                          & N@10   & 0.0468         & 0.0743            & \textbf{0.1125}       \\
\multicolumn{1}{c|}{}                          & N@50   & 0.0560          & 0.0843            & \textbf{0.1511}       \\ \midrule
\multicolumn{1}{c|}{\multirow{4}{*}{\sinspired}} & R@10   & 0.1436         & 0.2092            & \textbf{0.3417}       \\
\multicolumn{1}{c|}{}                          & R@50   & 0.2436         & 0.2704            & \textbf{0.5632}       \\
\multicolumn{1}{c|}{}                          & N@10   & 0.0927         & 0.1223            & \textbf{0.1772}       \\
\multicolumn{1}{c|}{}                          & N@50   & 0.1175         & 0.1344            & \textbf{0.2255}       \\ \midrule
\multicolumn{1}{c|}{\multirow{4}{*}{\sredial}}   & R@10   & 0.1670          & \textbf{0.2490}    & 0.2182                \\
\multicolumn{1}{c|}{}                          & R@50   & 0.2783         & 0.3347            & \textbf{0.4750}        \\
\multicolumn{1}{c|}{}                          & N@10   & 0.0937         & \textbf{0.1180}    & 0.1128                \\
\multicolumn{1}{c|}{}                          & N@50   & 0.1226         & 0.1370             & \textbf{0.1724}       \\ \bottomrule
\end{tabular}
\caption{Zero-shot performance of different LLMs compared to \aours. \texttt{4o-mini} and \texttt{4o} refer to GPT-4o-mini and GPT-4o, respectively.}
\label{tab:additional_llm_zs}
\end{table}

% \usepackage{booktabs}
% \usepackage{graphicx}
% \usepackage{array} % 상하 가운데 정렬을 위해 필요
\begin{table*}[ht!]\small
\centering
\begin{tabular}{>{\centering\arraybackslash}m{2cm}|m{13cm}} % 첫 번째 열 가운데 정렬
\toprule
\textbf{Stage} & \multicolumn{1}{c}{\textbf{Prompts}} \\ \midrule % 칼럼명 가운데 정렬
\textit{Superficial preference extraction}  & 
Given a dialogue history between the User and the System, find all aspects related to the movies the user currently seeks. Also, you must classify the preferences about each aspect, Like and Dislike. If there is nothing to mention about likes or dislikes, simply write "None." under the corresponding tag.
\newline
Dialogue history: \{\textbf{\textit{dialogHistory}}\} 
\newline
Response:
\newline
{[}Like{]} \{\{keyphrases or descriptions separated by comma\}\}  \newline
{[}Dislike{]} \{\{keyphrases or descriptions separated by comma\}\} \\ \midrule
\textit{Potential preference augmentation} & 
You are an advanced user's profile generator. Based on the conversation and the user's like/dislike preferences, use your reasoning to infer and expand upon the user's potential preferences. Augment key phrases related to the user's likes and dislikes, including preferences they may not have explicitly stated, to better guide personalized suggestions. If no explicit user preferences are provided, infer them from the conversation. Do not include any unrelated information; only state the user's preferences.
\newline \newline
User's preferences:  
\newline
\{\textbf{\textit{extractedPreferences}}\}
\newline
Conversation:
\newline
\{\textbf{\textit{dialogHistory}}\}
\newline \newline
Now, let's get started!  
\newline
{[}Like{]} \{\{Expanded keyphrases describing the user's likes\}\}  
\newline
{[}Dislike{]} \{\{Expanded keyphrases describing the user's dislikes\}\} \\
\bottomrule
\end{tabular}
\caption{Prompts for contrasting preference augmentation. Both \textbf{\textit{dialogHistory}} and \textbf{\textit{extractedPreferences}} are placeholders.}
\label{tab:appendix_augmentation_prompt}
\end{table*}

% \usepackage{booktabs}
% \usepackage{graphicx}
% \usepackage{array} % 상하 가운데 정렬을 위해 필요
\begin{table*}[ht!]\small
\centering
\begin{tabular}{>{\centering\arraybackslash}m{2cm}|m{13cm}} % 첫 번째 열 가운데 정렬
\toprule
\textbf{Stage} & \multicolumn{1}{c}{\textbf{Prompts}} \\ \midrule % 칼럼명 가운데 정렬
\textit{Review summarization}  & 
Given some popular reviews about Kids \{\textbf{\textit{title}}\}, describe what people liked and disliked about the movie under [Like] and [Dislike], respectively. If there is nothing to mention about like/dislike, simply write "None." under the corresponding tag.
\newline \newline
Here are some basic information about the movie and reviews about it:
\newline
Title: \{\textbf{\textit{title}}\}\newline
Genre: \{\textbf{\textit{genres}}\}\newline
Cast: \{\textbf{\textit{cast}}\}\newline
Director: \{\textbf{\textit{director}}\}
Reviews: \{\textit{\textbf{reviews}}\} \\ \midrule
\textit{Keyphrases generation} & 
Below are the common [Like] and [Dislike] from users about the \{\textbf{\textit{title}}\}. Based on this information, generate 5-8 keyphrases that represent user preferences and intentions for this movie, separated by commas. Do not include any other explanations or sentences.\newline\newline
Here is some basic information about the movie and users' preferences information:\newline
\{\textbf{\textit{userInformation}}\}\newline
The output format must strictly adhere to the following:\newline
[Like] {keyphrases or descriptions separated by comma}\newline
[Dislike] {keyphrases or descriptions separated by comma} \\
\bottomrule
\end{tabular}
\caption{Prompts for review summarization and keyphrases generation. Both \textbf{\textit{title}} and \textbf{\textit{userInformation}} are placeholders.}
\label{tab:appendix_item_prompt}
\end{table*}

\newpage

\appendix
\section{Further Study}
% Table~\ref{tab:zs_ablation}은 NV-Embed를 학습하지 않은 상태에서 query로 들어가는 input information을 ablation한 결과이다. 세 개의 벤치마크 데이터셋에 대해 모두 $C$, $L$, $D$를 모두 사용하는 것이 큰 성능 개선을 가져왔으며, 일관적으로 $C$와 $L$, $C$와 $D$를 사용한 경우에도 $C$만 사용한 경우보다 성능이 개선되었다. (i) 이는 우리가 증강한 데이터가 고품질의 데이터이며, 추천에 도움이 되는 데이터임을 확인할 수 있다. (ii) 또한, 우리가 제안한 서로 따로따로 scoring 방식이 사용자의 의도에 부합하는 방식임을 확인할 수 있다.

\subsection{Effect of Preference Ablation on Zero-shot Performance}
\label{sec:appendix_zero_shot_ablation}
Table~\ref{tab:zs_ablation} shows the results of ablating the input information used as a query without training NV-Embed. \textit{Avg. Gain} refers to the average performance gain when additional preferences are added compared to using only $T_c$. For the item-side textual representation, we concatenated $T_m$ with ${T_r}^{\mathrm{key}}$. Across all three benchmark datasets, using all of $T_c$, $T_{\mathit{l}}^{\mathrm{aug}}$, and $T_{\mathit{d}}^{\mathrm{aug}}$ resulted in significant performance improvements. Additionally, consistently using either $T_c$ and $T_{\mathit{l}}^{\mathrm{aug}}$ or $T_c$ and $T_{\mathit{d}}^{\mathrm{aug}}$ also outperformed using only $T_c$. (i) This demonstrates that the augmented data is of high quality and contributes positively to recommendation performance. (ii) Furthermore, it confirms that the proposed separate scoring method aligns well with the user's intent.

\subsection{Ablation Study on Small LM}
\label{sec:appendix_slm_ablation}
Table~\ref{tab:slm_ablation} presents the results of an ablation study on BERT, a small language model with 110M parameters. Notably, the highest performance is achieved when all preference components are incorporated. The ablation study further confirms that the components of \ours\ contribute consistently to performance improvement, even when applied to small language models.

\subsection{Effect of LLM in Contrasting Preference Expansion}
\label{sec:appendix_other_llm}
Table~\ref{tab:appendix_other_llm} shows the results of utilizing different LLMs in Contrasting Preference Expansion.
$\text{{C{\small ORAL}}}_{\text{ Mistral}}$ and $\text{{C{\small ORAL}}}_{\text{ gpt-4o-mini}}$ are variants that utilize \texttt{Mistral-7B-Instruct-v0.2} and \texttt{gpt-4o-mini} as LLMs in the Contrasting Preference Expansion step, respectively, and $\text{{C{\small ORAL}}}_{\text{ w/o } L, D, R}$ is a variant that does not use an expanded preference (\ie,  $T_{\mathit{l}}^{\mathrm{aug}}$, $T_{\mathit{d}}^{\mathrm{aug}}$, $T_{\mathit{r}}^{\mathrm{key}}$).
Both variants utilizing different LLMs outperform UniCRS and ECR and generally achieve better performance than the variant without expanded preferences.
These results highlight the effectiveness of \ours\ and further validate its scalability and generalizability.

\subsection{Zero-shot Performance of Different LLMs compared to \ours}
Table~\ref{tab:additional_llm_zs} shows the performance comparison between \ours\ and large language models \texttt{GPT-4o-mini} and \texttt{GPT-4o}. We used \texttt{gpt-4o-mini-2024-07-18} for \texttt{GPT-4o-mini} and \texttt{gpt-4o-2024-08-06} for \texttt{GPT-4o}, with the latter being a larger model than \texttt{GPT-4o-mini}. The experimental results demonstrate that \ours\ outperforms existing large language models on both \pearl\ and \inspired, and in the case of \redial, it significantly surpasses the LLMs in R@50 and N@50 metrics.

\section{Contrasting Preference Expansion}
\label{sec:appendix_prompt}
\subsection{User-side Expansion}
\begin{figure*}[t]
\centering
\includegraphics[width=1\linewidth]{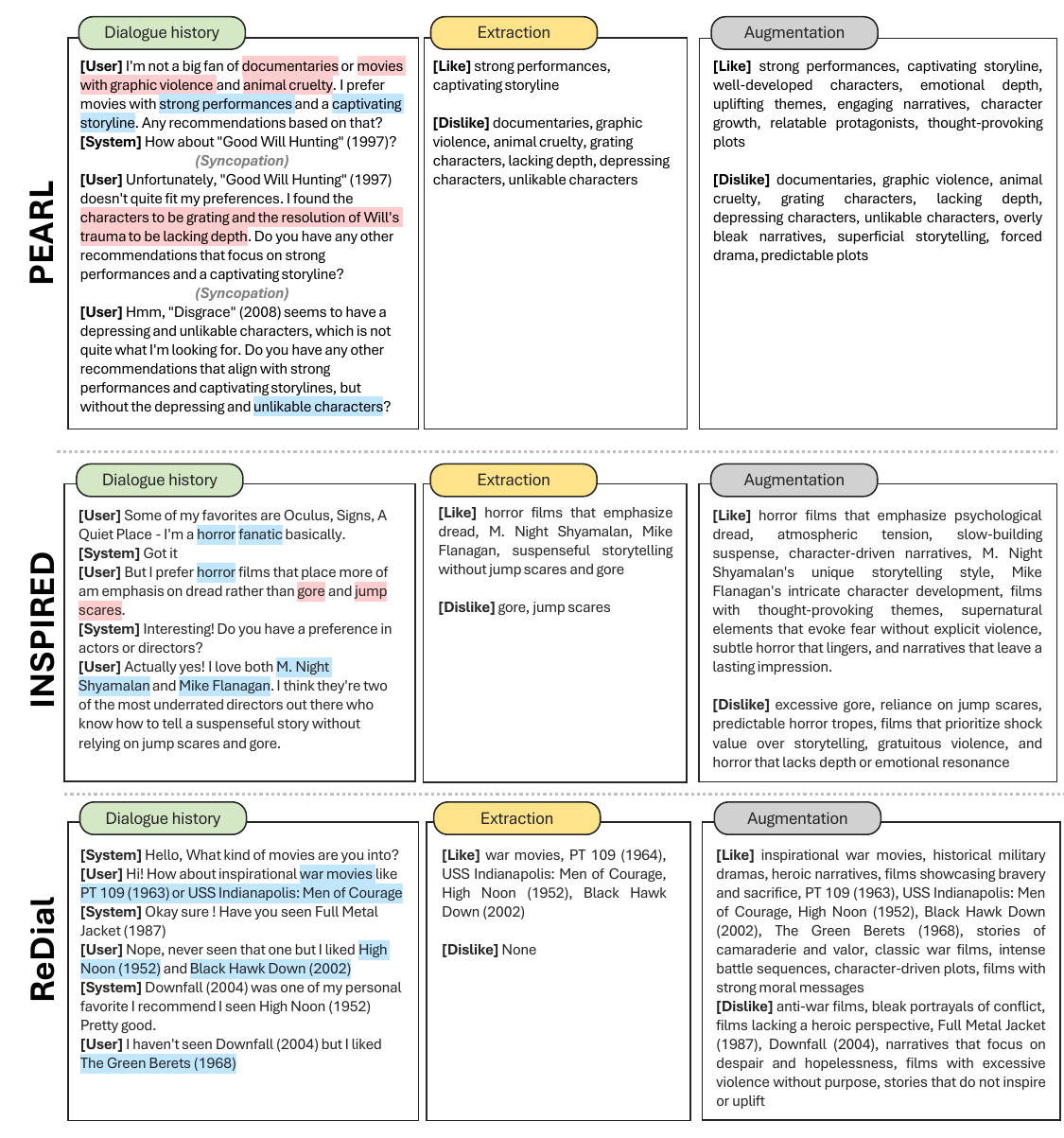} 
\caption{Step-by-step results applying the proposed contrasting preference expansion method to the \pearl, \inspired, and \redial\ datasets. "Extraction" corresponds to the superficial preference extraction phase, and "Augmentation" refers to the potential preference augmentation phase.}
\label{fig:appenddix_augmentation_examples}
\end{figure*}

% 1. Extraction 및 Augmentation에 사용된 prompt \\
% 2. 실제 extracted된 preference와 augmented된 preference의 예시 \\
% 3. 간단한 statistics (길이가 얼마나 늘어났는지, keyword가 얼마나 많아졌는지 (null 값 없어지는 효과)
% User-side expansion을 위해 사용한 prompt는 Table~\ref{tab:ambivalent preference augmentation prompt}와 같다. 
% % Table~\ref{tab:pref_prompt}는 제안 방법의 preference expansion 단계에서 사용되는 prompt를 나타내며, 
% 최종 결과의 예시는 Figure~\ref{fig:appenddix_augmentation_examples}에 나타나 있다.우리는 제안 방법을 통해 다양한 preference density를 갖는 벤치마크 데이터셋에서 extraction 단계에서 augmentation 단계로 갈 때, null values가 얼만큼 완화되었는지 Table~\ref{tab:null}에서 확인할 수 있다. 
The prompt used for \textit{user-side expansion} is shown in Table~\ref{tab:appendix_augmentation_prompt}. Figure~\ref{fig:appenddix_augmentation_examples} shows an example of the final results. 
Our augmentation method can enhance user preferences through reasoning power, even in cases where user preferences are scarcely revealed in the conversation. Experimentally, using the augmented potential preferences shows better performance than not using them, as demonstrated in Table~\ref{tab:Ablation study}. The datasets will be available upon acceptance.
% 우리의 방식으로 증강한 데이터는 대화 내 사용자의 선호가 거의 나타나지 않는 경우에도 reasoning power를 통해 사용자 선호를 증강할 수 있으며, 실험적으로 augmentated potential preference를 사용하는 것이 그렇지 않은 경우보다 더 나은 performance를 보인다(Table~\ref{tab:Ablation study}).
% Table~\ref{tab:} shows how the proposed method alleviates null values when transitioning from the extraction phase to the augmentation phase across benchmark datasets with varying preference densities.

\begin{figure}[t]\small
\centering
\includegraphics[width=1.0\linewidth]{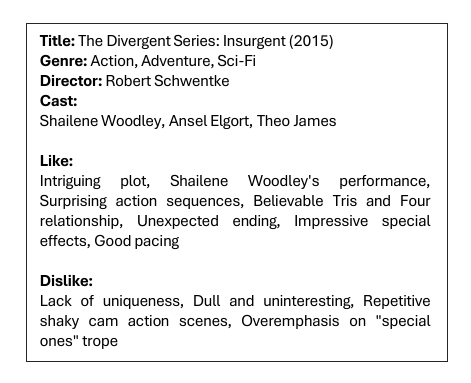} 
\caption{The examples of item-side expansion in Section~\ref{sec:contrasting_preference_expansion}}
\label{fig:appenddix_review_summary}
\end{figure}

\subsection{Item-side Expansion}
% 1. Item review 추출을 위한 prompt \\
% 2. 실제 description과 keyphrase의 예시
% 3. Case study (Extraction --> Augmentation 예시, 결과 아님)
% Item-side expansion을 위해서는 Section~\ref{sec:ambivalent_preference_expansion}에서 언급한 것처럼 크롤링한 리뷰로부터 리뷰 서머리를 추출하고, 이를 keyphrase로 바꾸는 과정이 필요하다. 해당 과정에서 사용한 prompt는 Figure~\ref{}와 같이 사용자들이 공통적으로 해당 아이템에 대해 가지는 common like/dislike preference 정보를 중심으로 요약한다. 그런 다음, keyphrase 형태로 나타내는데, Figure~\ref{fig:appenddix_review_summary}는 결과 예시 몇개를 보여준다.
For item-side expansion, as mentioned in Section~\ref{sec:contrasting_preference_expansion}, it is necessary to extract review summaries from crawled reviews and convert them into keyphrases. The prompt used in this process focuses on summarizing the common like/dislike preference information that users commonly associate with the item. The prompt used for \textit{item-side expansion} is shown in Table~\ref{tab:appendix_item_prompt}. These summaries are then transformed into keyphrases. Figure~\ref{fig:appenddix_review_summary} provides several examples of the results.

\section{Implementation Details}
\label{sec:appendix_implementation_baseline}
\noindent
\textbf{Traditional CRS models.}
Unlike \inspired~and \redial, \pearl~does not provide the knowledge graph traditional CRS requires. Therefore, we used DBpedia Spotlight~\cite{imple2011DBpediaSpotlight} to extract entities from the dialogue, and movie entities were constructed using the movie entities from \inspired~and~\redial.
While RevCore~\cite{pref/crs2021revcore} utilizes review data crawled from IMDb, we used our own crawled review data to ensure a fair comparison in our experiments.
ECR~\cite{pref/crs2024ecr} requires emotion labels. For \redial, we used the provided labels, whereas for \pearl~and~\inspired, which lack emotion labels, we inferred them using an LLM. Table~\ref{tab:ecr_prompt} details the prompt used for this inference. Additionally, since \pearl~and~\inspired~do not incorporate user feedback, we applied uniform weights to all items for feedback-aware item reweighting in ECR.
% \inspired와 \redial과는 다르게 \pearl은 knowledge graph를 제공하지 않는다.
% 따라서 우리는 DBpedia spotlight\cite{imple2011DBpediaSpotlight}를 사용하여 entity를 추출하였으며, 영화 entity들은 \inspired와 \redial의 movie entity들을 활용하여 구축하였다.
% RevCore~\cite{pref/crs2021revcore}는 IMDb에서 크롤링한 리뷰 데이터를 활용하지만, 공정한 비교를 위해 우리 실험에서는 우리가 크롤링한 리뷰 데이터를 활용한다.
% ECR~\cite{pref/crs2024ecr}은 emotion label을 필요로 한다. \redial은 제공해준 label을 사용했으며, emotion label이 없은 \pearl과 \inspired에 대해서는 LLM을 사용해서 inference했다. Prompt는 Table~\ref{tab:ecr_prompt}에 명시되어 있다. 또한 \pearl과 \inspired는 user feedback이 없으므로 모든 item에 동일한 weight를 적용하였다.

\vspace{0.5mm}g
\noindent
\textbf{LLM-based CRS models.}
We utilize ChatGPT\footnote{https://openai.com/} (\texttt{gpt-4o-mini}) as the backbone of LLM-based CRS.
It generates item titles based on user dialogue without any task-specific fine-tuning. 
Tables~\ref{tab:appendix_llm_zs_prompt}~and~\ref{tab:appendix_chatcrs_prompt} present the prompts used for the LLM-based approach.
We computed the average performance across two types of prompts for evaluation, as used in \cite{llm/crs2023llmzs, llm/crs2024chatcrs}.
% Table~\ref{tab:llm_prompt}는 LLM-based에서 사용된 프롬프트이다. 우리는 \cite{llm/crs2023llmzs, llm/crs2024chatcrs}에서 사용한 두 종류의 프롬프트에 대해 평가한  평균을 내었다. 

\vspace{0.5mm}
\noindent
\textbf{Retrieval-based CRS models.}
We implemented BM25~\cite{ir94BM25} using Pyserini~\cite{bm252021pyserini}, and DPR~\cite{ir2020dpr} was implemented with the BERT-base model~\cite{lm2019bert}.

\begin{table*}[] \small
\centering
\begin{tabular}{p{0.9\textwidth}} % 본문은 왼쪽 정렬, 칼럼명만 가운데 정렬
\toprule
\multicolumn{1}{c}{\textbf{Prompt}} \\ \midrule % 칼럼명을 가운데 정렬
You are an expert in emotion analysis. Given a target user's dialogue utterance and the dialogue history of the target user's dialogue utterance, identify the emotions expressed in the target user's dialogue utterance from the provided options. The options are as follows: a. like b. curious c. happy d. grateful e. negative f. neutral g. nostalgia h. agreement i. surprise. \newline Output only the corresponding letter, and nothing else. Note that you only need to analyze the emotions in the target user's dialogue utterance, not the dialogue history. \newline Dialogue history: \{\textbf{\textit{dialogHistory}}\} \newline Target user dialogue utterance: \{\textbf{\textit{utternace}}\} \\ \bottomrule
\end{tabular}
\caption{Prompts for emotion classifier used to reproduce ECR. Both \textbf{\textit{dialogHistory}} and \textbf{\textit{utternace}}  are placeholders.}\label{tab:ecr_prompt}
\end{table*}

% \usepackage{booktabs}
% \usepackage{graphicx}
% \usepackage{array} % 상하 가운데 정렬을 위해 필요
\begin{table*}[ht!]\small
\centering
\begin{tabular}{>{\centering\arraybackslash}m{2cm}|m{13cm}} % 첫 번째 열 가운데 정렬
\toprule
\textbf{Reference} & \multicolumn{1}{c}{\textbf{Prompts}} \\ \midrule % 칼럼명 가운데 정렬
Zero-shot~\cite{llm/crs2023llmzs}  & 
Pretend you are a movie recommender system.
\newline
I will give you a conversation between a user and you (a recommender system). Based on the conversation, you reply me with 50 recommendations without extra sentences.
\newline
The format of the recommendation list is: no. title (year).
\newline
Here is the conversation:
\newline
\{\textbf{\textit{dialogHistory}}\} \\ \midrule
ChatCRS~\cite{llm/crs2024chatcrs} & 
You are an excellent conversational recommender that helps the user achieve recommendation-related goals through conversations. Given the dialogue history, your task is to generate appropriate item recommendations for the dialogue.
\newline
You reply me with 50 recommendations without extra sentences. The format of the recommendation list is: no. title (year).
\newline
Dialogue history:
\newline
\{\textbf{\textit{dialogHistory}}\} \\
\bottomrule
\end{tabular}
\caption{Prompts used for LLM zero-shot recommendation. Adapted from the prompts used for movie recommendation in \cite{llm/crs2023llmzs, llm/crs2024chatcrs}. We reported the average performance of these two prompts. Both \textbf{\textit{dialogHistory}} and \textbf{\textit{extractedPreferences}} are placeholders.}
\label{tab:appendix_llm_zs_prompt}
\end{table*}

% \usepackage{booktabs}
% \usepackage{graphicx}
% \usepackage{array} % 상하 가운데 정렬을 위해 필요
\begin{table*}[ht!]\small
\centering
\begin{tabular}{>{\centering\arraybackslash}m{2cm}|m{13cm}} % 첫 번째 열 가운데 정렬
\toprule
\textbf{Reference} & \multicolumn{1}{c}{\textbf{Prompts}} \\ \midrule % 칼럼명 가운데 정렬
Zero-shot~\cite{llm/crs2023llmzs}  & 

% Start
Pretend you are a movie recommender system. I will give you a conversation between a user and you (a recommender system), along with the related knowledge triplets. Based on this information, you reply to me with 50 recommendations without extra sentences. The format of the recommendation list is: no. title (year).
\newline
Here is the conversation:\{\textbf{\textit{dialogHistory}}\}
\newline
Here is the related knowledge triplets: [\{\textbf{\textit{knowledgeTriplets}}\}]
% End
\\ \midrule
ChatCRS~\cite{llm/crs2024chatcrs} & 
% start
You are an excellent conversational recommender that helps the user achieve recommendation-related goals through conversations. Given the dialogue history and a knowledge triplets, your task is to genenrate appropriate item recommendations for the dialogue. You reply me with 50 recommendations without extra sentences. The format of the recommendation list is: no. title (year).
\newline
Dialogue history: \{\textbf{\textit{dialogHistory}}\}
\newline
Knowledge triplets: [\{\textbf{\textit{knowledgeTriplets}}\}]
% End
\\ \bottomrule
\end{tabular}
\caption{Prompts used for ChatCRS~\cite{llm/crs2024chatcrs} recommendation. Adapted from the prompts used for movie recommendation in \cite{llm/crs2023llmzs, llm/crs2024chatcrs}. We reported the average performance of these two prompts. Both \textbf{\textit{dialogHistory}} and \textbf{\textit{extractedPreferences}} are placeholders.}
\label{tab:appendix_chatcrs_prompt}
\end{table*}

\end{document}